\documentclass[trackchanges,twocolumn]{aastex7}

\usepackage{graphicx}
\usepackage{enumitem}
\usepackage{amsmath}
\usepackage{pifont}
\usepackage{CJK}
\usepackage{soul}
\usepackage{makecell}

\newcommand{\cmark}{\ding{51}}%
\usepackage{tikz}

\newcommand{\mgas}[0]{M_{\rm gas}}
\newcommand{\mdust}[0]{M_{\rm dust}}

\newcommand{\coo}[0]{\rm C^{18}O}
\newcommand{\twelveco}[0]{\rm {}^{12}CO}
\newcommand{\thirteenco}[0]{\rm {}^{13}CO}
\newcommand{\sigmadust}[0]{\Sigma_{\rm dust}}
\newcommand{\sigmagas}[0]{\Sigma_{\rm gas}}

\newcommand{\um}[0]{{\rm\mu  m}}

\shorttitle{DiskMINT: IM~Lup}
\shortauthors{D. Deng et al.}

\graphicspath{{./}{figures/}}

\received{June 24, 2025}
\revised{September 02, 2025}
\accepted{September 18, 2025}

\submitjournal{ApJ}

\begin{document}

\title{DiskMINT: Self-Consistent Thermochemical Disk Models with Radially Varying Gas and Dust -- Application to the Massive, CO-Rich Disk of IM~Lup}

\author[0000-0003-0777-7392]{Dingshan Deng} 
\affiliation{Lunar and Planetary Laboratory, the University of Arizona, Tucson, AZ 85721, USA}
\email{dingshandeng@arizona.edu}

\correspondingauthor{Dingshan Deng}
\email{dingshandeng@arizona.edu}


\author[0000-0002-3311-5918]{Uma Gorti}
\affiliation{NASA Ames Research Center, Moffett Field, CA 94035, USA}
\affiliation{Carl Sagan Center, SETI Institute, Mountain View, CA 94043, USA}
\email{ugorti@seti.org}

\author[0000-0001-7962-1683]{Ilaria Pascucci}
\affiliation{Lunar and Planetary Laboratory, the University of Arizona, Tucson, AZ 85721, USA}
\email{pascucci@arizona.edu}

\author[0000-0003-0522-5789]{Maxime Ruaud}
\affiliation{Carl Sagan Center, SETI Institute, Mountain View, CA 94043, USA}
\email{maxime.ruaud@gmail.com}

\begin{abstract}

Disks around young stars are the birthplaces of planets, and the spatial distribution of their gas and dust masses is critical for understanding where and what types of planets can form.
We present self-consistent thermochemical disk models built with \texttt{DiskMINT}, which extends its initial framework to allow for spatially decoupled gas and dust distributions. 
\texttt{DiskMINT} calculates the gas temperature based on thermal equilibrium with dust grains, solves vertical gas hydrostatic equilibrium, and includes key processes for the CO chemistry, specifically selective photodissociation, and freeze-out with conversion \textbf{of} CO/CO$_2$ ice.
We apply \texttt{DiskMINT} to study the IM~Lup disk, a large, massive disk, yet with an inferred CO depletion of up to 100 based on earlier thermochemical models.
By fitting the multi-wavelength SED along with the millimeter continuum, ${\rm C^{18}O}$ radial emission profiles, we find $0.02-0.08\,{\rm M_\odot}$ for the gas disk mass, which \textbf{is} consistent with the dynamical-based mass within the uncertainties.
We further compare the derived surface densities for dust and gas and find that the outer disk is drift-dominated, with a dust-to-gas mass ratio of approximately 0.01-0.02, which is likely insufficient to meet the conditions for the streaming instability to occur.
Our results suggest that when interpreted with self-consistent thermochemical models, ${\rm C^{18}O}$ alone can serve as a reliable tracer of both the total gas mass and its radial distribution. 
This approach enables gas mass estimates in lower-mass disks, where dynamical constraints are not available, and in fainter systems where rare species like ${\rm N_2H^+}$ are too weak to detect.

\end{abstract}



\section{Introduction}
\label{sec:intro}

Planets form in disks around young stars, and one of the most fundamental properties of a protoplanetary disk is its total mass and how that mass is distributed.
The gas content sets the upper limit for forming gas giants, while the dust content determines the building blocks of planetary cores. 
Moreover, the dust-to-gas mass ratio and its radial variation are key diagnostics of planetesimals and planet formation efficiency, gas disk dispersal, and overall disk evolution \citep[see, e.g.,][for a recent review]{Miotello_PPVII_2023}. 
Accurately estimating both gas and dust masses, and how they are distributed, is essential for constraining disk physics; however, obtaining reliable estimates remains a complex challenge.

The dust disk mass ($\mdust$) and its radial distribution are typically estimated from the dust thermal emission at (sub)millimeter wavelengths, which is usually assumed to be optically thin \citep[e.g.,][]{Ansdell_lupus_2016, Pascucci_mass_2016}.
Recent work \citep[e.g.][]{Carasco-Gonzalez_HLTau_2016, Zhu_dust_scattering_2019, xin_measuringMdust_2023} found that dust emission is not always optically thin, even at longer millimeter wavelengths ($\gtrsim 3\,{\rm mm}$).
Estimates can be improved by fitting the multi-wavelength spectral energy distribution (SED) with a radiative transfer model \citep[e.g., ][]{woitke_diana_2019, rilinger_determiningMdust_2023}.
However, $\mdust$ estimates still rely on dust opacity, which depends on the composition of dust grains and their size distribution.

The gas disk mass ($\mgas$) and its radial distribution can be estimated similarly from optically thin emission lines, but this procedure requires modeling the thermal structure and chemical processes in the disk. 
Isotopologues of abundant species are expected to be optically thin and can therefore be used as potential tracers of disk mass. 
$\mathrm{HD}$, the isotopologue of the most abundant molecule $\mathrm{H_2}$, is favored for measuring $\mgas$ (however, see \citealt{Ruaud2022} for its limitations), but has been detected only in three disks \citep{Bergin_HD_2013, McClure_HD_2016} and is currently inaccessible.
Carbon monoxide (${\rm CO}$) is the second most abundant molecule that spatially co-exists with $\mathrm{H_2}$, and the emission from its rarer isotopologues, such as $\thirteenco$ and $\coo$ are believed to be the most promising tracers of the gas content in disks.
Observations from the Atacama Large Millimeter/submillimeter Array (ALMA), however, have been contentious. 
Many thermochemical models tend to produce more CO isotopologue emission than what is observed.  
While converting the model predictions and data into gas masses, some argue for low gas-to-dust mass ratios in these disks and low $\mgas$ \citep[e.g.,][]{Ansdell_lupus_2016, Miotello2016, miotello_lupus_2017, Long_Cham_2017}, while others argue -- based on higher $\mgas$ estimates from $\mathrm{HD}$ \citep{Bergin_HD_2013, McClure_HD_2016} and dynamical masses \citep[e.g.,][]{lodato_dynamical_mass_2023, martire_rotation_2024} -- that the disk gas phase $\mathrm{CO}$ is reduced as disks evolve, possibly due to preferential sequestration of CO into planetesimals due to dynamical processes \citep[e.g., ][]{Bergin_n_Williams_mass_2017, Bosman_n_Banzatti_twhya_2019, Krijt_COdep_2020, Powell_COdep_2022}. 
Some spatially resolved ALMA observations and thermochemical modeling have led to inferred global reductions of up to two orders of magnitude of $\mathrm{CO}$ in Myr-old disks \citep[e.g.,][]{law_IMLup_molecules_vertical_2021, zhang_MAPS_molecules_2021}.

More recently, in a series of works \citep[][]{Ruaud2022, pascucci_noCOdep_2023, Deng_2023_diskmint, ruaud_cold_2024}, it has been argued that with self-consistent thermochemical disk models, there is no need for such global reductions in gas phase CO. 
Three processes were identified as critical to interpreting emission from the rarer CO isotopologues: (1) freeze-out with treatment of grain-surface chemistry where subsequent ${\rm CO}$/${\rm CO_2}$ ice conversion is the main reaction; (2) CO isotopologue selective photodissociation; and (3) self-consistent gas disk density and temperature structure coupled by the vertical pressure equilibrium. 
These models are also applied to the large disks in the Lupus star-forming region in analyzing the $\mathrm{[CI]}$ line emission \citep[][]{pascucci_noCOdep_2023} as well as to the latest ALMA AGE-PRO observations that cover more typical Lupus disks \citep[][]{Deng_2025_AGEPRO_III_Lupus}.
Both works demonstrate that there is no need to globally reduce the gas-phase $\mathrm{CO}$ abundance to reproduce the measured CO and C line fluxes. 
Their derived $\mgas$ are orders of magnitude larger than earlier $\mathrm{CO}$-based $\mgas$ estimates, and more in line with typical $\mathrm{HD}$- and dynamical-based mass estimates.
This finding has also been supported by \citet{zwicky_dancing_2025} through another independent thermochemical model framework.

Motivated by the above, in \citet{Deng_2023_diskmint}, we presented \texttt{DiskMINT}\footnote{\url{https://github.com/DingshanDeng/DiskMINT}} (Disk Model for INdividual Targets), which is a self-consistent model based on \citet{Ruaud2022} but simplified by assuming the gas and dust are thermally well-coupled everywhere and by using a reduced chemical network that still includes all necessary processes related to the CO chemistry.
We applied \texttt{DiskMINT} to the disk around RU~Lup, a young high-accreting star \citep[][]{alcala_x-shooter_2017} whose disk was earlier inferred to have only a $\mgas \sim 1.5 \times 10^{-3}\,M_\odot$ from CO isotopologue emission \citep[][]{miotello_lupus_2017}. 
We fit the long wavelength ($\gtrsim 100\,\um$) continuum emission to derive its $\mdust$ and thermal structure and the optically thin $\coo$ line emission, as well as the velocity and radial intensity profiles, to find a higher $\mgas \sim 1-2 \times 10^{-2}\,M_\odot$ without assuming arbitrary CO depletion factors.

Here, we expand and improve \texttt{DiskMINT} with new functions that enable us to decouple the dust and gas: the radial surface densities of gas and dust can be independently varied, and \texttt{DiskMINT} computes self-consistent vertical structure from hydrostatic equilibrium with dust settling.
Therefore, we can infer not just total masses but also the radial surface density distributions from high-resolution ALMA observations of dust and gas. 
These new features of \texttt{DiskMINT} enable us to investigate the evolution of the disk.
We demonstrate this method on the large disk of IM~Lup (Section~\ref{sec:observation}), which has extensive observations \citep[e.g.,][]{cleeves_IMLup_coupled_2016, huang_disk_2018, law_IMLup_molecules_vertical_2021, zhang_MAPS_molecules_2021}.
In Section~\ref{sec:models}, we describe details of the model with its new functions and present the main results. 
A discussion of the results and their implications is provided in Section~\ref{sec:discussions}, followed by a summary in Section~\ref{sec:summary}.

\section{The Large Disk of IM~Lup} 
\label{sec:observation}


IM~Lup (Sz~82) is a $\sim 1$\,Myr-old star \citep[][]{mawet_direct_2012} located at a distance of $158.9\,{\rm pc}$ (\texttt{Gaia} DR3; \citealt{gaia_collaboration_gaia_2023}) in the Lupus star-forming region. 
Its stellar mass is approximately solar $\sim 1.1\,M_\odot$ \citep[][]{teague_maps_2021}, and thus serves as a young analog to the Sun.

\begin{figure*}[ht]
\gridline{\fig{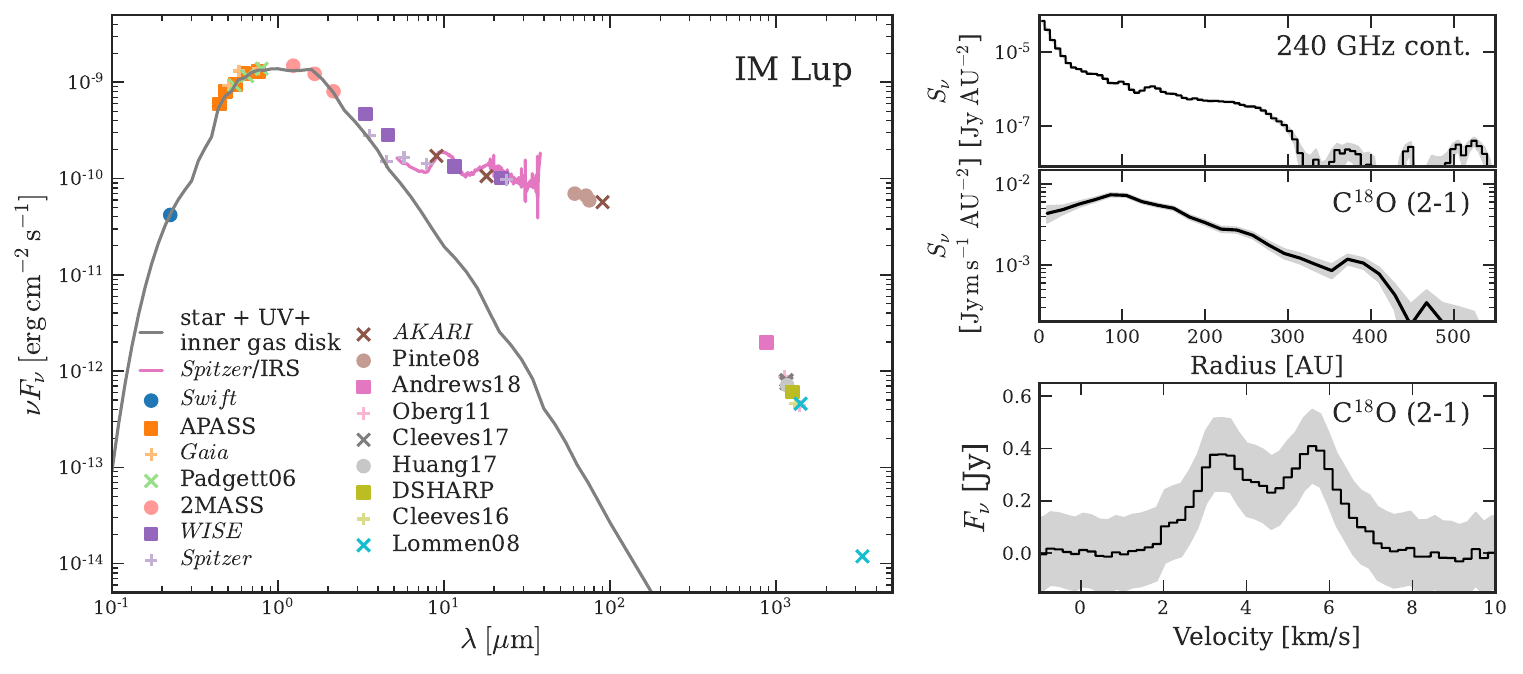}{0.95\textwidth}{}}
\caption{Observational data that is used in this work.
    Left panel: spectral energy distribution (SED). Photometry in colored markers and \textit{Spitzer/}IRS spectrum \citep[][]{evans_molecular_2003, evans_spitzer_2009} in magenta. 
    A combined \texttt{BT-settl} stellar photosphere spectrum \citep{Allard_2003IAUS..211..325A, Allard_2011ASPC..448...91A} with a 8,500 K blackbody radiation fitting the UV data-point, and another blackbody radiation with 1,500 K from the radius of $R_{\star}$ to 0.055 AU represents the accreting gas disk at the inner edge, is shown as a gray line. 
    Right panels: ALMA Band~6 continuum (240 GHz) radial profile from DSHARP \citep[][]{andrews_disk_2018}, and $\mathrm{C^{18}O}$ (2-1) radial profile (lower right) and disk-integrated line profile (lower left) from MAPS \citep[][]{oberg_molecules_2021}. 
    The disk-integrated line profile is compiled from \texttt{GoFish} \footnote{\url{https://github.com/richteague/gofish}} \citep[][]{teague_gofish_2019}, and the radial profiles are measured at each radius with an elliptical annulus with the inclination and position angles in the continuum image and the $\coo$ moment-0 maps, respectively.
    SED references: UV photometry from \emph{Swift} \citep[][]{cleeves_IMLup_coupled_2016}; optical photometry from APASS \citep[][]{henden_vizier_2016}, \emph{Gaia} \citep[][]{gaia_collaboration_gaia_2023}, \citet{padgett_spitzer_2006}; infrared photometry from 2MASS \citep[][]{skrutskie_two_2006}, \emph{WISE} \citep[][]{wright_wide-field_2010}, \emph{Spitzer} \citep[][]{padgett_spitzer_2006}, \emph{AKARI} \citep[][]{ishihara_akariirc_2010}; and (sub)millimeter data from \citet{pinte_probing_2008}, \citet{andrews_disk_2018}, \citet{oberg_disk_2011}, \citet{cleeves_variable_2017}, \citet{huang_alma_2017}, DSHARP \citep[][]{huang_disk_2018}, \citet{cleeves_IMLup_coupled_2016}, and \citet{lommen_sma_2008}. 
    }
\label{fig:obs_SED_n_C18O}
\end{figure*}

IM~Lup has been extensively observed at multiple wavelengths from UV to radio, resulting in the multi-wavelength SED shown in Figure~\ref{fig:obs_SED_n_C18O}, where average photometry is reported for multi-epoch observations.
Additionally, deep high-angular resolution ALMA observations have spatially resolved its disk, both in the dust and gas, including in multiple CO isotopologues \citep[e.g., ][]{van_kempen_searching_2007, cleeves_IMLup_coupled_2016}, and IM~Lup was chosen as a target by the ALMA large programs DSHARP \citep[][]{andrews_disk_2018} and MAPS \citep[][]{oberg_molecules_2021}.
Interestingly, IM~Lup was found to have the largest dust and gas disk sizes among all disks studied in both programs. 
Its dust disk, measured as the radius containing 90\% of the flux, is $R^{90}_{\rm mm} \sim 264\pm1\,{\rm AU}$ \citep[][]{huang_disk_2018}. 
Its $\twelveco$ gas disk is nearly double the size of the dust disk $R^{90}_{\twelveco}\sim 481\pm6\,{\rm AU}$ \citep[][]{law_IMLup_molecules_vertical_2021}.
Additionally, these ALMA observations also found the disk to have multiple dust rings. 

Large masses have been inferred for the disk of IM~Lup.
A large $\mdust \sim 1.0-2.0\times10^{-3}\,M_\odot$ was inferred from its bright (sub)millimeter continuum emission observed from SMA and ALMA using multiple wavelengths.
Assuming a gas-to-dust mass ratio of 100, its gas mass is inferred to be $\mgas \sim 0.1-0.2\,M_\odot$, which is $\sim 10-20\%$ of its stellar mass \citep[][]{pinte_probing_2008, cleeves_IMLup_coupled_2016, zhang_MAPS_molecules_2021}.
This large $\mgas$ is independently supported by deviations from the Keplerian rotation of CO isotopologue lines induced by the disk self-gravity \citep[][]{lodato_dynamical_mass_2023, martire_rotation_2024}.
The derived high disk mass is in line with other kinematic signatures, such as spiral arms, suggesting that this disk could be on the verge of gravitational instability.

However, the $\mgas$ inferred from thermochemical models matching CO isotopologue data is reported to be only $\sim 2.0\times10^{-3}\,M_\odot$ \citep{zhang_MAPS_molecules_2021}. 
The authors can reconcile models with data if gas-phase CO is homogeneously reduced by a factor of 100 from their model output.
They also found a significant variation in the required CO abundance depletion factor in the radial profile.

The young age, large radial extent, and large inferred CO abundance anomaly make the disk around IM~Lup an interesting target to attempt to model in a self-consistent fashion. 
Motivated by our earlier thermochemical modeling results \citep[][]{Deng_2023_diskmint}, we use \texttt{DiskMINT} to investigate the mass and surface density distribution of the IM~Lup disk. 
Using a minimal set of input parameters including stellar and disk properties (see Table~\ref{Tab:star_n_disk}), we aim to match the spatially resolved dust and gas observations to infer the evolutionary status of this young disk.

\begin{deluxetable}{lccc}[h]
\tablecaption{The stellar and disk properties of IM~Lup \label{Tab:star_n_disk}}
\tablewidth{0.5\textwidth}
\tablehead{
\colhead{Parameter} & \colhead{Symbol} & \colhead{Value} & \colhead{Reference}
}
\startdata
    \textit{Stellar Properties} &&& \\
    stellar mass & $M_\mathrm{\star}$  &  1.1$\,M_\odot$ & [1]\\
    stellar radius & $r_\mathrm{\star}$ & 2.5$\,R_\odot$ & [1]\\
    distance & $D$ & 158$\pm$0.5 pc & [2] \\
    extinction & $A_V$ & 0.7 mag & [3] \\
     mass accretion rate & $\dot{M}_\mathrm{acc}$ & $1\times 10^{-8}\,M_\odot\,\mathrm{yr^{-1}}$ & [4] \\
    \hline
    \textit{Disk Properties} &&& \\
    inclination angle & $i$ & $47.5\pm0.3$ & [1, 5] \\
    position angle & ${\rm PA}$ & $144.5\pm0.5$ & [1, 5]\\
    source velocity & $v_\mathrm{sr}$ & $4.5\,{\rm km\,s^{-1}}$ &  [1]\\ 
    dust disk size & $R^{90}_{\rm dust}$ & $264\pm1\,{\rm AU}$ &  [5]\\
    gas disk size & $R^{90}_{\rm gas}$ & $481\pm6\,{\rm AU}$ & [6]
\enddata
\tablecomments{References: [1] \citet{zhang_MAPS_molecules_2021}; [2] \citet{gaia_collaboration_gaia_2023}; [3] \citet{gunther_disk-bearing_2010}; [4] \citet{alcala_x-shooter_2017}; [5] \citet{huang_disk_2018}; [6] \citet{law_IMLup_molecules_vertical_2021}.}
\end{deluxetable}

\section{Models and Main Results}
\label{sec:models}

\texttt{DiskMINT} is an open-source tool built on \texttt{RADMC-3D} \citep[][]{Dullemond_radmc-3d_2012} for radiative transfer, and it includes a reduced chemical network \citep[][]{Ruaud2022} to determine the $\coo$ emission from dust and gas disk models.
\texttt{DiskMINT} computes self-consistent disk structures, iterating between density and temperature, and has a detailed treatment of CO isotopologue chemistry, including isotope-selective photodissociation and grain-surface chemistry with ${\rm CO}$/${\rm CO_2}$ ice conversion.

In this work, we utilize two different approaches to investigate the gas and dust distributions in the disk around IM~Lup.
We start with a well-mixed model, following the same approach used on the disk around RU~Lup as in \citet{Deng_2023_diskmint}.
In this model, we focus on fitting the integrated flux densities and luminosities for both dust continuum and gas line emission.
We fit the long wavelengths ($\gtrsim 100\,\um$) SED to derive the dust disk mass ($\mdust$) and thermal structure.
Next, we fit the integrated $\coo$ luminosity ($L_{\coo}$) to estimate the gas disk mass ($\mgas$).
The details for this well-mixed model are summarized in Section~\ref{subsec:simple_model_A}.

In the next approach, we use high angular-resolution ALMA observations for IM~Lup that provide additional constraints on the dust and gas distributions ($\sigmadust$ and $\sigmagas$).
In this data-driven approach, we iteratively solve for $\sigmadust$ and $\sigmagas$ by matching observations and present the detailed steps for finding the surface densities and construct a structured model in Section~\ref{subsec:data_driven_model_B}.

\begin{figure*}
\gridline{\fig{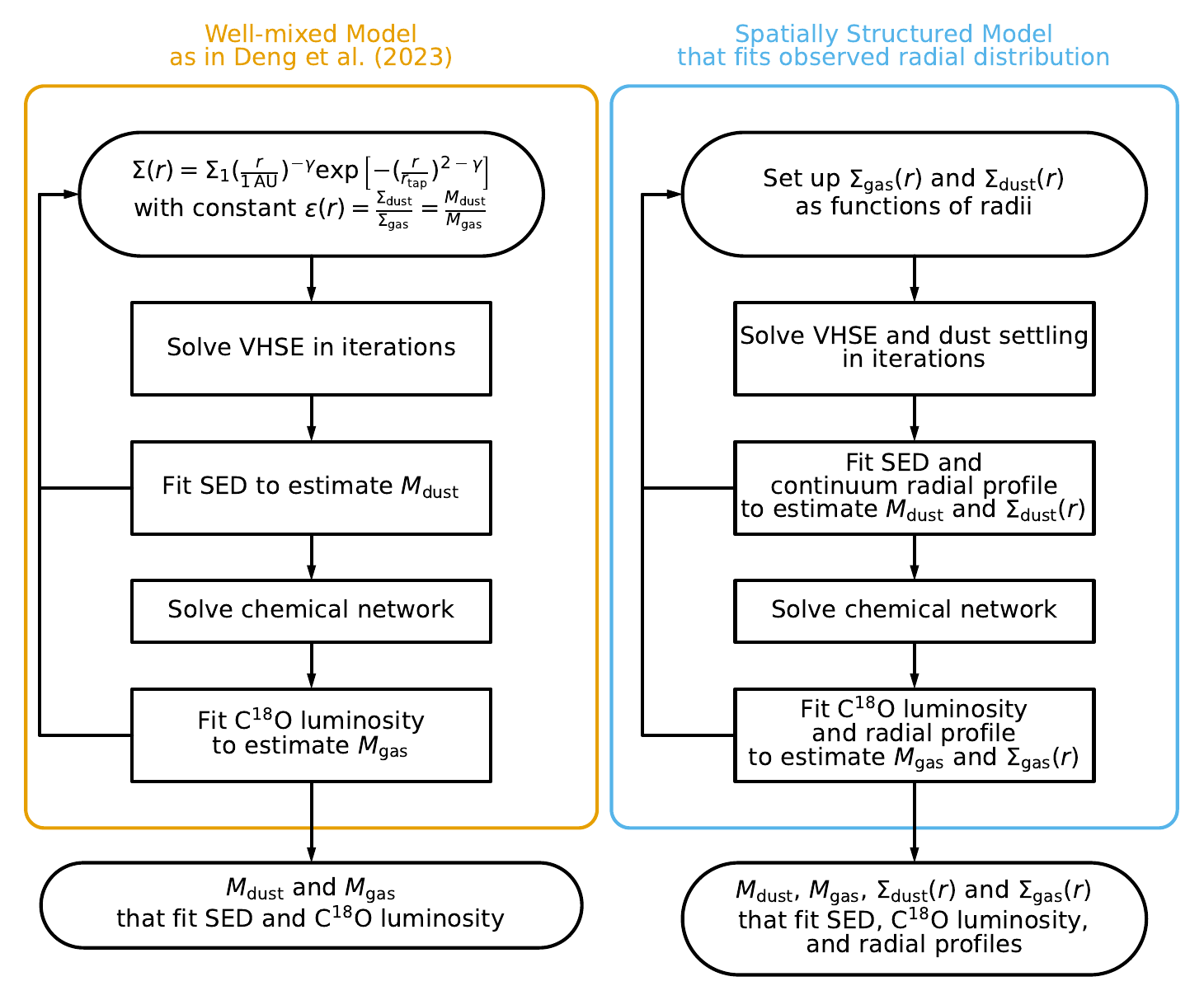}{0.80\textwidth}{}}
\caption{Flow chart summarizing the two approaches to estimate disk masses. The main difference between them is the surface density distributions of gas and dust. We solve vertical hydrostatic equilibrium (VHSE) for all models and include dust settling for the structured model (see Sections~\ref{subsec:data_driven_model_B} and ~\ref{subsec:theoretical_model_C}), and we use the reduced chemical network that considers the isotope-selective photodissociation and grain surface chemistry to get the CO abundances. This flow chart only shows the main steps in \texttt{DiskMINT}, and more details are described in \citet{Deng_2023_diskmint}.}
\label{fig:flow_chart}
\end{figure*}

In addition, we attempt to gain a more theoretical understanding of the implications of the derived $\sigmagas$ on disk evolution from the structured model, and test whether the dust and gas surface densities are consistent with what is expected from theory. 
Details are summarized in Section~\ref{subsec:theoretical_model_C}.
The two main approaches are also summarized in a flow chart (Figure~\ref{fig:flow_chart}), and the results are presented in subsections below. 

For all of the models, we adopt an input stellar spectrum with three components: a \texttt{BT-settl} stellar photosphere spectrum with $T_{\rm eff} = 4300\,{\rm K}$ \citep[][]{Allard_2003IAUS..211..325A, Allard_2011ASPC..448...91A}; a $8500\,{\rm K}$ blackbody radiation fitting the UV data-point in the SED; and a $1500\,{\rm K}$ blackbody radiation at $0.055\,{\rm AU}$ representing the accreting gas in the inner disk.
After obtaining the self-consistent disk density structure in hydrostatic equilibrium, we run the reduced chemical network using the ISM chemical abundances \citep[][]{Ruaud2022, Deng_2023_diskmint} to determine the CO abundances.
Then, we utilize \texttt{LIME} \footnote{\url{https://github.com/lime-rt/lime}} \citep[Line Modelling Engine;][]{BrichHogerheijde_LIME_2010} to generate synthetic images.
Finally, radial profiles generated from these synthetic images are compared to the observed radial profiles.

\subsection{Well-mixed model: 
Perfectly coupled gas and dust with a uniform dust-to-gas mass ratio}
\label{subsec:simple_model_A}

We first build a well-mixed model where the dust and gas have a fixed mass ratio all through the disk (at all $(r,z)$) using the same method as in \citet{Deng_2023_diskmint}, and we summarize model inputs and procedures here. 
All the adopted and inferred parameters are summarized in Table~\ref{Tab:model_A_para}.

\begin{deluxetable}{lcccc}
\tablecaption{Parameters for well-mixed model \label{Tab:model_A_para}}
\tabletypesize{\normalem}     
\setlength{\tabcolsep}{0.5pt}       
\tablewidth{0.45\textwidth}
\tablehead{
\colhead{Parameter} & \multicolumn{2}{c}{Symbol} & \multicolumn{2}{c}{Value}
}
\startdata
    \textit{Dust Properties} &&&& \\
     Composition &&& DIANA standard dust\tablenotemark{a} \\
    minimum size & \multicolumn{2}{c}{$a_\mathrm{min}$}  &  \multicolumn{2}{c}{$1\times10^{-5}$ cm} \\
    maximum size\tablenotemark{b} & \multicolumn{2}{c}{$a_\mathrm{max}$}  &  \multicolumn{2}{c}{$0.06\,\mathrm{cm}$} \\
    exponential slope\tablenotemark{b} & \multicolumn{2}{c}{$q$}  &  \multicolumn{2}{c}{3.2} \\ 
    \hline
    \textit{Vertical Structure} &&&& \\
    vertical density distribution &&& \multicolumn{2}{c}{\makecell[c]{solve VHSE in iterations \\ dust is well-mixed and \\ well-coupled with gas}} \\
    \hline
    \textit{Radial Structure} &&&& \\
    inner radius of the disk & \multicolumn{2}{c}{$r_\mathrm{in}$}  &  \multicolumn{2}{c}{0.055 AU}\\
    gas surface density & \multicolumn{2}{c}{$\sigmagas(r)$} & \multicolumn{2}{c}{adopt Equation~\ref{eq:power-law-sigma}} \\
    $-$ characteristic radius & \multicolumn{2}{c}{$r_c$}  &  \multicolumn{2}{c}{100 AU} \\
    $-$ surface density slope & \multicolumn{2}{c}{$\gamma$} & \multicolumn{2}{c}{1.0} \\
    dust surface density & \multicolumn{2}{c}{$\sigmadust(r)$} & \multicolumn{2}{c}{\makecell[c]{well-mixed and \\ well-coupled with gas}} \\
\enddata
\tablenotetext{a}{The DIANA standard dust \citep[][]{Woitke_diana_2016, min_multiwavelength_2016} is a mixture of pyroxene and carbon in a mass ratio of 0.87:0.13 and with a porosity of 25\% (see Section~\ref{subsec:simple_model_A}).}
\tablenotetext{b}{$a_{\rm max}$ and $q$ are inferred from SED fitting.}
\end{deluxetable}

\paragraph{Dust properties}
The dust number density follows a power-law distribution as 
\begin{equation}\label{eq:power-law-size}
    n(a) \propto a^{-q};\ (a_{\rm min} \leq a \leq a_{\rm max})
\end{equation}
where $a$ is the radius of the dust grains and $q$ is the exponent describing the size distribution.
We fix $a_{\rm min} = 10^{-5}\,\mathrm{cm}$ and fit the SED to infer the $a_{\rm max}$ and $q$. 
We use the \texttt{optool}\footnote{\url{https://github.com/cdominik/optool}} \citep[][]{dominik_optool_2021} package to compute the wavelength-dependent dust opacity using the Bruggeman \citep[][]{bruggeman_berechnung_1935} mixing rule, and we use the DIANA standard opacities \citep[][]{Woitke_diana_2016, min_multiwavelength_2016}, which are a mixture 60\% of amorphous silicates (\citealt{dorschner_steps_1995}, ${\rm Mg_{0.7}Fe_{0.3}SiO_{3}}$) with 15\% amorphous carbon (\citealt{zubko_optical_1996}, BE-sample), with 25\% porosity, by volume. 
The DIANA standard opacities are chosen as they are similar to the grains in the ISM, and are more representative of smaller grains at the $\coo$ emitting surface.
We divide the dust grains into 20 different size bins -- which is sufficient to constrain the resulting thermal structure to an accuracy of $5\%$ (for details, see Section~2.1 in \citealt{Deng_2023_diskmint}) -- in logarithmic-scale from $a_{\rm min}$ to $a_{\rm max}$, and the opacities are calculated for grains in each size bin, to be used in \texttt{DiskMINT} to obtain the thermal structure and estimate the dust continuum emission.

\paragraph{A self-similar surface density distribution}
This well-mixed model uses the surface density distribution that follows the similarity solutions of viscously evolving disks \citep{lynden-bell_evolution_1974, Hartmann1998}:
\begin{equation}\label{eq:power-law-sigma}
    \Sigma(r) = \Sigma_1 (\frac{r}{1\,\mathrm{AU}})^{-\gamma} \exp{\left[- (\frac{r}{r_{c}})^{2-\gamma}\right]};    (r_{\mathrm{in}} < r< r_{\mathrm{out}})
\end{equation}
where the $\Sigma_{1}$ is the surface density at $1\,{\rm AU}$, which is determined by the input dust disk mass $\mdust$, and $\gamma$ is the power-law index for this surface density distribution.
We adopt typical values for the characteristic radius of $r_{c} = 100\,\mathrm{AU}$ and $\gamma = 1.0$ as in previous thermochemical models \citep[e.g.,][]{Miotello2014, Miotello2016, miotello_lupus_2017, Ruaud2022, trapman_exoalma_2025, Trapman_AGEPRO_V_gas_masses}. 
This approach is used in \texttt{DiskMINT} to estimate  $\mgas$ from $\coo$ for disks that do not have high-resolution observations.
We set the inner disk radius $r_{\rm in} = 0.055\,\mathrm{AU}$ where the dust sublimates as the dust temperature goes to $T_{\rm dust}(r_{\rm in}) \sim 1500\,{\rm K}$.
We also adopt $r_{\rm out} \sim 1000\,\mathrm{AU}$ to ensure that all disk emission is included beyond $r_c$.

\paragraph{dust-to-gas mass ratio}
We adopt a constant dust-to-gas mass ratio ($\varepsilon$) everywhere in the disk which is the defining feature of the well-mixed model, and therefore the total gas disk mass is simply related to the total dust disk mass, 
\begin{equation}
     \frac{\mdust}{\mgas} = \frac{\sigmadust(r)}{\sigmagas(r)} = \varepsilon(r) = {\rm Constant}.
\end{equation}

\begin{figure}[ht]
\gridline{\fig{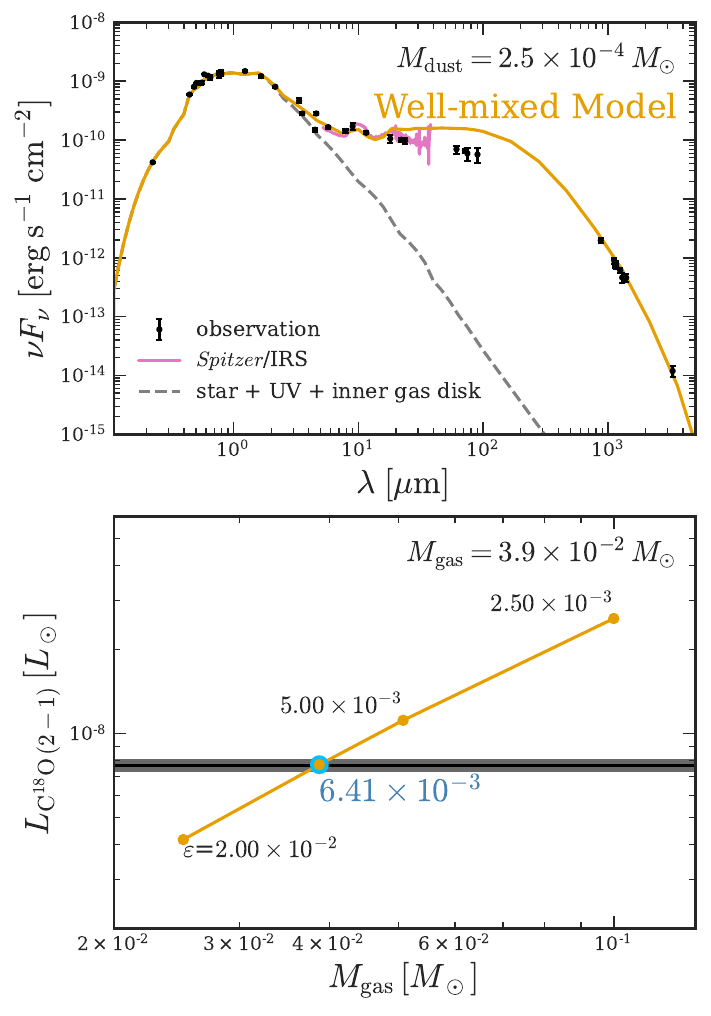}{0.45\textwidth}{}}
\caption{The synthetic SED compared with observation (top) and the synthetic $L_{\coo(2-1)}$ for different $\mgas$ compared with observation (bottom) for the well-mixed model. In the top panel, the observations are shown in black points with errors, the \textit{Spitzer/}IRS spectrum is shown in the magenta line, and the input stellar spectra with the stellar photosphere, UV, as well as the inner gas disk are shown in the dashed gray line. The synthetic SED of the model is shown in orange line, with the inferred $\mdust$ shown on the upper right corner. In the bottom panel, the black horizontal line shows the $\coo(2-1)$ observation from MAPS, with its uncertainties shown as the dark gray area. The blue points show the model points with different $\varepsilon$, and the best-fit model with $\varepsilon=6.41\times10^{-3}$ is circled in blue, with its corresponding $\mgas$ shown on the upper right corner.}
\label{fig:Result_Model_A}
\end{figure}

Following \citet{Deng_2023_diskmint}, we run \texttt{DiskMINT} by solving for disk vertical hydrostatic equilibrium.
Firstly, we utilize \texttt{DiskMINT} to fit the SED at long wavelengths $\lambda \gtrsim 100\,\um$ (Figure~\ref{fig:Result_Model_A}, top panel), where the disk emission is likely optically thin and a better indicator of dust mass.
The SED is generated using \texttt{RADMC3D} with the disk inclination ($i = 47.5\arcdeg$) from ALMA continuum images by DSHARP \citep[][]{huang_disk_2018}.
The inferred dust disk and structural parameters are reported in Table~\ref{Tab:model_A_para}.
We estimate $\mdust$ from the SED, mainly referring to the $\mathrm{mm}$-wavelength continuum flux densities from ALMA, and we find $\mdust = 2.5\times10^{-4}\,M_\odot = 83\,M_\oplus$. 
This estimate is about a factor of $\sim 10$ smaller than the $\mdust$ reported in some works \citep[e.g.,][]{zhang_MAPS_molecules_2021}, mainly due to the differences in opacities, see Section~\ref{subsec:discuss_dust_mass} for details.

After the dust disk parameters including $\mdust$ are determined, we run \texttt{DiskMINT} with different constant $\varepsilon$ around the fiducial ISM value of $0.01$, including $\varepsilon = 0.02$, $0.01$, and $0.005$, where we obtain the estimated thermal structure and chemical abundances of gas including $\coo$. 
Then we utilize \texttt{LIME} to conduct non-Local Thermal Equilibrium line radiative transfer as in \citet{Deng_2023_diskmint} and generate synthetic images using the same imaging parameters from MAPS data, with a synthetic beam size of $0.15\arcsec \times 0.15\arcsec$ and field-of-view out to 20\arcsec. 
The adopted gas disk inclination ($i = 47.5\arcdeg$), position angles (${\rm PA} = 144.5\arcdeg$), and the systematic velocity ($v_{\rm sys} = 4.5 \,\mathrm{km\,s^{-1}}$) are also from MAPS CO isotopologue data \citep[][]{zhang_MAPS_molecules_2021}.
Then we compare the synthetic $\coo$ luminosity with observations (Figure~\ref{fig:Result_Model_A}, bottom panel) and find the best-fit $\varepsilon=6.41\times10^{-3}$ (the point with blue circle), corresponding to $\mgas = 3.9\times10^{-2}\,M_\odot$.



\subsection{Spatially Structured Model: decoupled gas and dust and fitting the radial profiles}
\label{subsec:data_driven_model_B}

IM~Lup has high angular resolution ALMA observations on both dust continuum and $\coo$ line emission, which enables us to radially resolve its dust and gas line emission.
The well-mixed model above only considers the integrated fluxes over the entire disk and implicitly assumes that the gas and dust are well-coupled everywhere in the disk; however, this is an oversimplification. 
Gas and dust are in fact observed to have different distributions \citep[e.g.,][]{villenave_observations_2020}, but the radial distributions of neither the dust nor $\coo$ gas line emission of the well-mixed model well reproduce the observations as shown in Appendix~\ref{appendix:model_A_to_B}.

Here, we add new features to \texttt{DiskMINT} to create a more realistic structured model with spatially decoupled dust and gas distributions.
In the vertical direction, we consider decoupling of gas and dust due to settling and solve for this iteratively with the gas vertical distribution set by hydrostatic equilibrium.
In the radial direction, we make use of the spatially resolved emission maps from ALMA to derive $\sigmagas(r)$ and $\sigmadust(r)$.
The entire disk structure (both radial and vertical) is iteratively determined to match the observational constraints described below. 

\subsubsection{Vertical distribution: dust settling}
\label{subsec:dust_settling}

\begin{figure}
    \centering
    \includegraphics[width=0.99\linewidth]{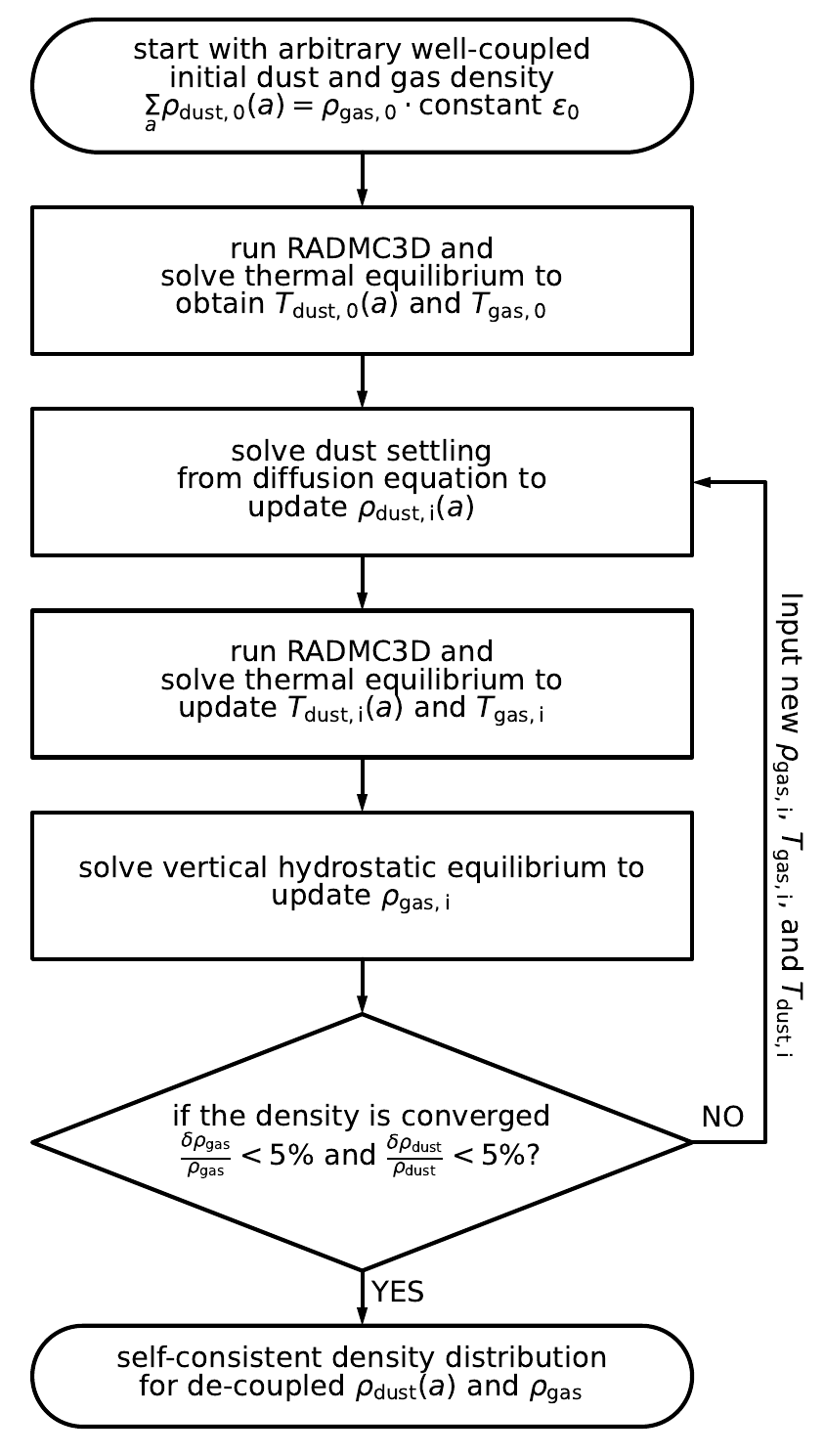}
    \caption{The flow chart demonstrates the steps to calculate dust settling when solving the vertical density distribution. The details on the procedures of solving thermal equilibrium and the vertical hydrostatic equilibrium are presented in \citet{Deng_2023_diskmint} Section~2.1 and Figure~1. For all models, the well-mixed case where dust is coupled with gas,  and the structured model where they decouple due to settling, the same convergence criterion is used --- that $\rho_{\rm gas}$ and $\rho_{\rm dust}$ attain a 5\% accuracy with iterations.}
    \label{fig:dust_settling_flow_chart}
\end{figure}

The extended and flared disk around IM~Lup also enables us to obtain information on the disk's vertical structure.
The dust vertical profile is determined by the balance between gravitational settling toward the disk midplane and upward lift due to turbulence \citep[e.g., ][]{dubrulle_dust_settling_1995, dullemond_flaring_2004}.  
We do not adopt the standard approach of calculating particle scale height \citep[e.g.,][]{dubrulle_dust_settling_1995, Estrada_2016_particle_distribution} as these expressions implicitly assume a vertically isothermal temperature profile.
We consider the transport of dust particles by the turbulent fluctuations as a diffusion process in steady state \citep{dullemond_flaring_2004, fromang_n_Nelson_settling_2009}:

\begin{equation}\label{eq:dust_diffusion_original}
    \frac{\partial}{\partial z}\left(\ln{\frac{\rho_{\rm dust}}{\rho_{\rm gas}}}\right) = -\frac{\Omega^2 \tau_{\rm s}}{D}z,
\end{equation}
where $\Omega$ is the Keplerian angular speed ($\Omega = \sqrt{G M_\star / r^3}$), $\tau_{\rm s}$ is the stopping time, $D$ is a diffusion coefficient that quantifies the turbulent diffusivity, and both $\rho_{\rm dust}$ and $\rho_{\rm gas}$ are functions of position $(r,z)$. 
The stopping time is a function of particle sizes and can be written as:
\begin{equation}
    \tau_{s} = \frac{\rho_\bullet a}{\rho_{\rm gas} c_{\rm s}},
\end{equation}
where $\rho_\bullet$ is the bulk density of dust grains, $a$ is the particle size, and $c_{\rm s}$ is the sound speed ($c_{\rm s} = \sqrt{k_{\rm B} T_{\rm gas}/\mu}$ where $\mu$ is the mean molecular weight). The diffusion coefficient can also be expressed as:
\begin{equation}
    D = \frac{\alpha_{\rm v} c_{\rm s} h}{{\rm Sc}},
\end{equation}
where $\alpha_{\rm v}$ is the viscous turbulent parameter, $h = c_{\rm s}/\Omega$ is the pressure scale height, and ${\rm Sc}$ is the Schmidt number assumed to be $\sim 1$ \citep[][]{fromang_n_Nelson_settling_2009}.
Therefore, Equation~\ref{eq:dust_diffusion_original} can be written as:
\begin{equation}\label{eq:dust_diffusion_full}
    \frac{\partial}{\partial z}(\ln{\frac{\rho_{\rm dust}(a)}{\rho_{\rm gas}}}) = -\frac{1}{\alpha_{\rm v}} \cdot \frac{\rho_\bullet }{\rho_{\rm gas}}  \cdot \frac{\Omega^3 z a}{ c_{\rm s}^3}.
\end{equation}
Here, $\rho_{\rm gas}$ and $T_{\rm gas}$ are computed in the models, and the dust vertical structure is essentially determined by the only free parameter $\alpha_{\rm v}$.
We solve this dust particle diffusion equation when solving the vertical hydrostatic equilibrium for gas \citep[see Section~2.1 in][]{Deng_2023_diskmint} in iterations. 
This updated procedure is described in a flow chart (Figure~\ref{fig:dust_settling_flow_chart}).
We first start with an arbitrary density distribution where gas and dust are well-coupled with the same dust size distribution everywhere in the disk, then we go through iterations including solving radiative transfer with \texttt{RADMC3D} to estimate $T_{\rm dust}(a)$, thermal equilibrium to estimate $T_{\rm gas}$, vertical hydrostatic equilibrium to estimate $\rho_{\rm gas}$, and the dust diffusive equation to estimate $\rho_{\rm dust}(a)$.
This results in the large grains settling towards the midplane while the small grains stay mostly at the disk surface. However, the overall dust size distribution at each radius and the vertically integrated (surface) densities for each size, $\sigmadust(a, r)$ -- are kept fixed. 
In the end, when the densities converge in these iterations, we obtain the self-consistent vertical density distribution resulting from dust settling and gas pressure equilibrium.

\subsubsection{Radial distribution}
\label{subsection:modify_sigma_for_model_B}

Besides the vertical profile, the dust and gas are also decoupled radially in the structured model.
Here, we set the gas and dust surface densities separately, both as functions of $r$, and therefore the dust-to-gas mass ratio is a function of $r$,
\begin{equation}
    \varepsilon(r) = \sigmadust(r)/\sigmagas(r). 
\end{equation}
We adjust $\sigmadust(r)$ and $\sigmagas(r)$ based on dust continuum and $\coo$ line observations, respectively, following the steps below.

\paragraph{Step 1. Derive dust surface density}
Because the dust thermal emission is expected to be optically thin in the majority of the outer disk, the intensity per area ($I_{A}$) at each radius is proportional to the dust surface density at each radius,
\begin{equation}
    I_{\nu, A}(r) = B_\nu (T_{\rm dust}) \kappa_{\nu}\sigmadust;\,({\rm when\ optically\ thin})
\end{equation}
where $B_\nu (T_{\rm dust})$ is the blackbody radiation with the temperature $T_{\rm dust}$, and $\kappa_{\nu}$ is the average opacity that is generated by \texttt{optool} with the adopted composition and grain size distribution.
We compare the radial profile inferred from the synthetic continuum image with the profile inferred from a high angular resolution image from DSHARP \citep[][]{huang_disk_2018}, and we find the best-fit $\sigmadust(r)$ in iterations.
We derive the continuum radial profile from the images using elliptical annuli, utilizing the inclination and position angles from the observation, and get the intensity per area for each annulus. 
In each iteration, we compare the radial profile from the model with observations, 
\begin{equation}
    R_{{\rm cont.}\,i}(r) = \frac{I_{A, {\rm observation}}(r)}{I_{A, {\rm model}\,i}(r)}
\end{equation}
where $I_{A, {\rm model}\,i}(r)$ is the continuum intensity per unit area at radius $r$ and model iteration $i$.
Then, we modify the surface density distributions based on the ratio between the two
\begin{equation}
\label{eq:fit_sigmadust}
    \Sigma_{{\rm dust,model\,}i+1}(r) =  \Sigma_{{\rm dust,model\,}i}(r) R_{{\rm cont},i}(r)
\end{equation}
We start from the image generated from the well-mixed model, and modify $\sigmadust$ until it matches the observed continuum radial intensity profile.

\paragraph{Step 2. Derive gas surface density}
Similarly, we derive $\sigmagas$ based on the radial profile of the $\coo$ line emission because it is optically thin and its emission layer is close to the disk midplane \citep[e.g.,][]{Miotello2014, Miotello2016, Ruaud2022, Deng_2023_diskmint}.
We also find best-fit $\sigmagas(r)$ in iterations,
\begin{equation}
\label{eq:fit_sigmagas}
    \Sigma_{{\rm gas,model\,}i+1}(r) =  \Sigma_{{\rm gas,model\,}i}(r) R_{{\rm \coo},i}(r)
\end{equation}
where $R_{{\rm \coo},i}(r)$ is the ratio between the observation and the synthetic image from the model $i$ for $\coo$ line radial profile.

We start from what is adopted in the well-mixed model (Section~\ref{subsec:simple_model_A}) for both dust and gas, and both $\sigmadust(r)$ and $\sigmagas(r)$ are set in iterations by fitting the dust continuum and $\coo\,(2-1)$ line radial profiles (the middle two panels). 
The radial profiles for both synthetic images and observations are created by measuring the total flux per area at each radius inside an annulus that is inclined to match the observed geometry, and they are presented in the flux per unit area ($\mathrm{Jy\,AU^{-2}}$ and $\mathrm{Jy\,m\,s^{-1}\,AU^{-2}}$), which should be proportional to the mass per unit area (i.e., surface densities) when the emission is optically thin.
It typically takes three iterations to find the converged best-fit $\sigmadust(r)$ and five iterations for the converged best-fit $\sigmagas(r)$, and we demonstrate this method by presenting some of the intermediate models in Appendix~\ref{appendix:model_A_to_B}. 

\begin{figure}[t]
\gridline{\fig{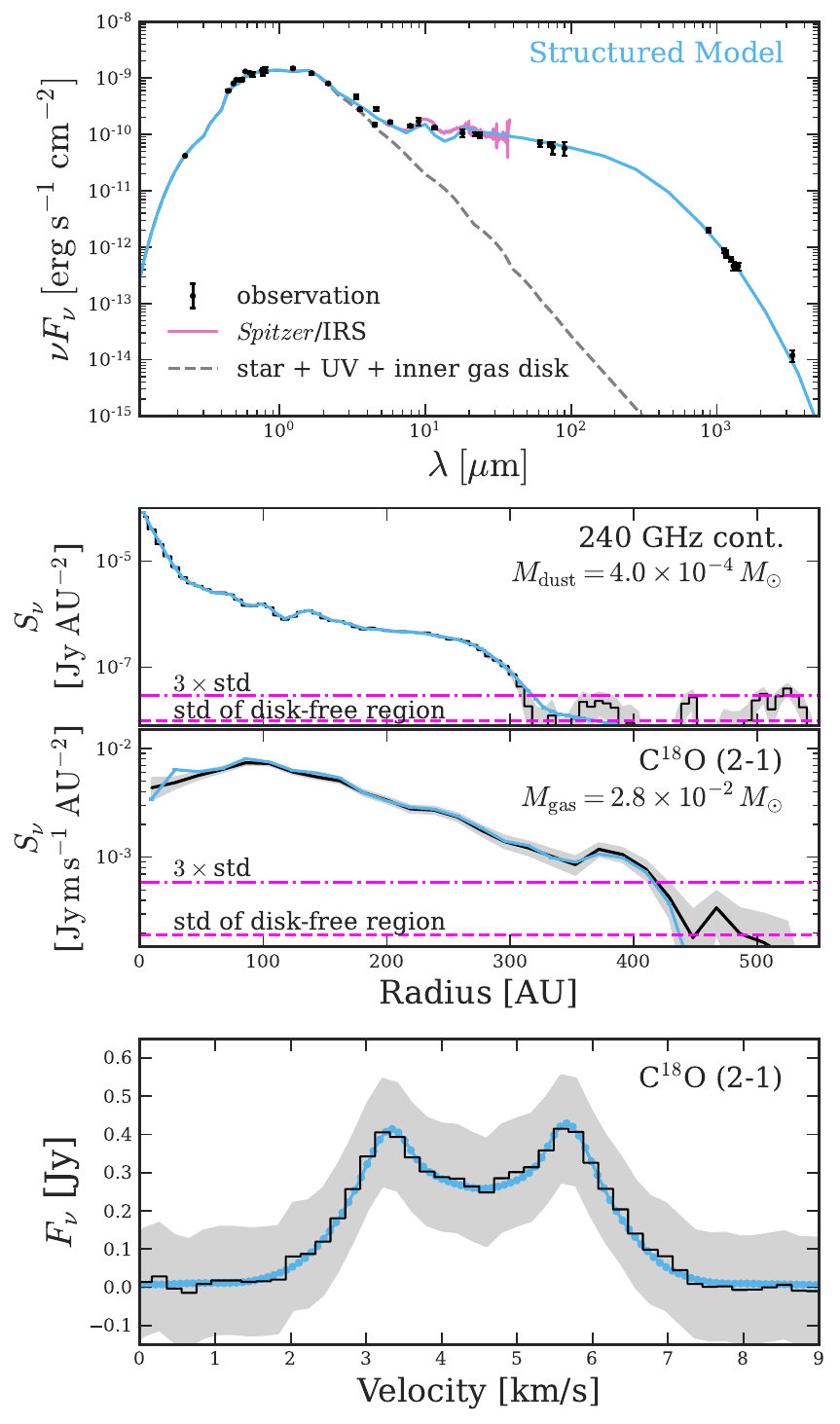}{0.45\textwidth}{}}
\caption{Compare observations and the results from structured model, in SED (top), cont. and $\coo(2-1)$ radial profiles (middle two panels) and $\coo(2-1)$ line spectra (bottom). The captions in the SED follow the top panel of Figure~\ref{fig:Result_Model_A}, with the synthetic SED from the structured model shown in blue. The continuum and $\coo(2-1)$ radial profiles are created by measuring the total flux per area at each radius inside an annulus that is tilted to match the observed geometry, with the best-fit $\mdust$ and $\mgas$ shown at the upper right corners. The line spectra are made with \texttt{GoFish} without de-projection. Observations are shown in black lines, along with their uncertainties, in the radial profile and the line spectra.}
\label{fig:Result_Model_B}
\end{figure}

\subsubsection{Results for Structured Model} 

We also summarize the parameters we adopt for the structured model in Table~\ref{Tab:model_B_para}, where we adopt the same dust composition, $a_{\rm min}$, and the $r_{\rm in}$ as the well-mixed model.
The main results including the SED fitting, continuum and $\coo(2-1)$ radial profile, and the $\coo(2-1)$ line spectra for the structured model are presented in Figure~\ref{fig:Result_Model_A} with the best-fit $\mdust = 4.0\times 10^{-4}\,M_\odot = 133\,M_\oplus$ and $\mgas =2.8\times 10^{-2}\,M_\odot$. 

\begin{deluxetable}{lcccc}[t]
\tablecaption{Parameters for Structured Model \label{Tab:model_B_para}}
\tabletypesize{\normalem}     
\setlength{\tabcolsep}{0.5pt}       
\tablewidth{0.45\textwidth}
\tablehead{
\colhead{Parameter} & \multicolumn{2}{c}{Symbol} & \multicolumn{2}{c}{Value}
}
\startdata
    \textit{Dust Properties} &&&& \\
    Composition &&& \multicolumn{2}{c}{DIANA standard dust} \\
    minimum size & \multicolumn{2}{c}{$a_\mathrm{min}$}  &  \multicolumn{2}{c}{$1\times10^{-5}$ cm} \\
    maximum size\tablenotemark{a} & \multicolumn{2}{c}{$a_\mathrm{max}$}  &  \multicolumn{2}{c}{$0.06\,\mathrm{cm}$} \\
    exponential slope\tablenotemark{a} & \multicolumn{2}{c}{$q$}  &  \multicolumn{2}{c}{3.2} \\ 
    \hline
    \textit{Vertical Structure} &&&& \\
    vertical density distribution &&& \multicolumn{2}{c}{\makecell[c]{solve VHSE with dust \\ settling in iterations}} \\
    $-$ turbulence\tablenotemark{a} & \multicolumn{2}{c}{$\alpha_{\rm v}$} & \multicolumn{2}{c}{$1\times10^{-3}$}\\
    \hline
    \textit{Radial Structure} &&&& \\
    inner radius of the disk & \multicolumn{2}{c}{$r_\mathrm{in}$}  &  \multicolumn{2}{c}{0.055 AU}\\
    gas surface density & \multicolumn{2}{c}{$\sigmagas(r)$} & \multicolumn{2}{c}{fit to $\coo\,(2-1)$ observation} \\
    dust surface density & \multicolumn{2}{c}{$\sigmadust(r)$} & \multicolumn{2}{c}{fit to continuum observation} \\
\enddata
\tablenotetext{a}{$a_{\rm max}$, $q$ and $\alpha_{\rm v}$ are inferred from SED fitting. We find the same values for $a_{\rm max}$, $q$ as the well-mixed model (Table~\ref{Tab:model_A_para}).}
\end{deluxetable}

We find this structured model with the same dust size distribution (described by $a_{\rm max}$ and $q$) as the well-mixed model can also fit the slope of the SED at longer wavelengths $\lambda > 100\,\um$, and the dust settling results in a higher $\mdust$ to fit the SED. 
Furthermore, this model improves the match to the  SED at $\lambda \sim 100\,\um$ by lowering disk flaring (compare Figure~\ref{fig:Result_Model_A} and Figure~\ref{fig:Result_Model_B} upper panels  and Appendix~\ref{appendix:settling_models} for details).
For the structured model, we also find the best-fit value for the turbulent parameter $\alpha_{\rm v} = 10^{-3}$ based on the SED fitting for (for details on how $\alpha_{\rm v}$ can affect the SED and the modeling, see Appendix~\ref{appendix:settling_models}).

\subsection{Placing the derived gas and dust surface densities in the context of dust dynamical evolution}
\label{subsec:theoretical_model_C}


In the structured model, $\sigmadust(r)$ and $\sigmagas(r)$ were separately set to match the observations in iterations and find a more extended $\sigmagas(r)$ compared to $\sigmadust(r)$, and the $\sigmadust(r)$ was truncated at $R_{\rm dust}$ to match the observations.
The observed difference in the radial extents of gas and dust may be due to different sensitivities to line and continuum emission, but it is also expected based on theoretical considerations, as gas and dust undergo distinct evolutionary processes. 
Dust is expected to drift radially inward relative to the gas, and the dust size distribution is believed to be set by collisional equilibrium between coagulation and fragmentation processes \citep[e.g.,][]{testi_dust_2014}.

Next, we construct another model where we adopt a gas surface density as in the structured model, and compute a dust surface density in accordance with current dust evolution theory to compare with the data. The purpose of this is to test the extent to which the gas and dust distributions of the disk around IM~Lup are consistent with theoretical expectations. 
Specifically, we aim to examine whether dust fragmentation and drift can distribute the dust surface density for different grain sizes ($\sigmadust(r,a)$) to result in the observed distributions.
We note that this testing is mainly to examine the extent of the gas and dust distribution in the context of dust evolutionary theories, and is not a separate attempt or approach to infer the surface densities.

\begin{figure}
\gridline{\fig{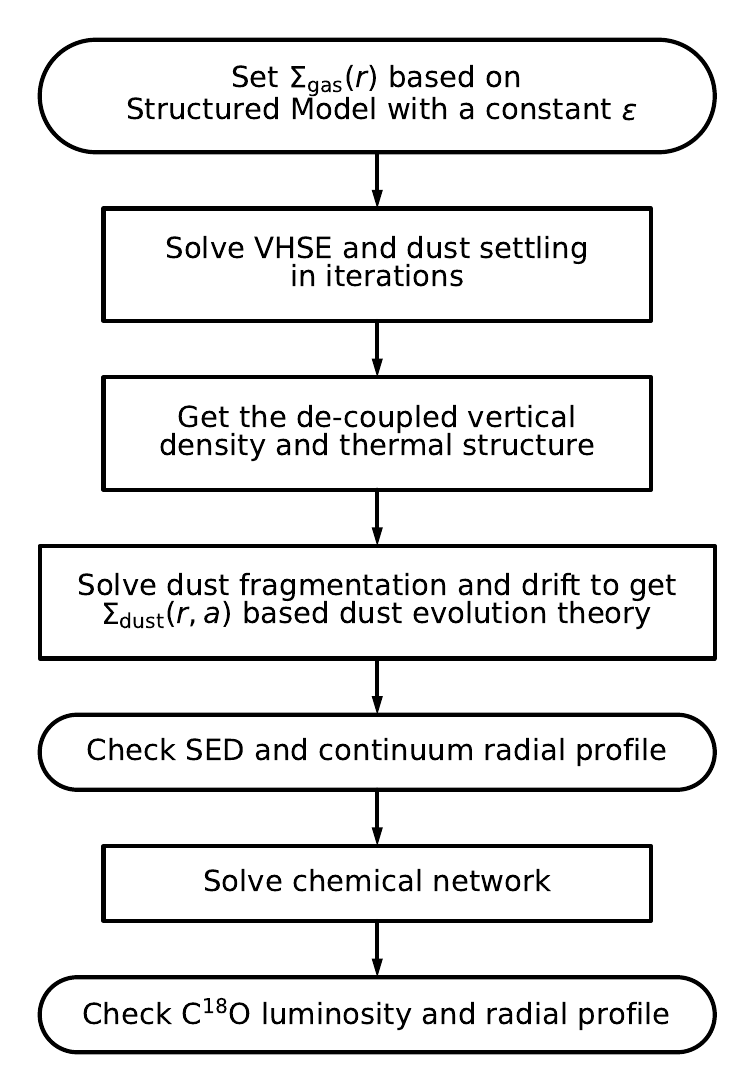}{0.49\textwidth}{}}
\caption{Flow chart summarizing the steps for the theoretical test of dust dynamics.}
\label{fig:flow_chart_C}
\end{figure}

The flowchart summarizing this procedure is presented in Figure~\ref{fig:flow_chart_C}, and the details of the redistribution of the dust and how $\sigmadust(r,a)$ are obtained are presented in the Appendix~\ref{appendix:model_C_theories}.
Here, we present its setup and results.

\subsubsection{Setup for the dust dynamics test}
\label{subsection:model_C_setup}

The $\sigmagas(r)$ is based on the results from the structured model, but we use instead a fit to the \citet{lynden-bell_evolution_1974} self-similar solution for surface density (Equation~\ref{eq:power-law-sigma}) as is often assumed in theoretical work.
We fit the $\sigmagas(r)$ profile that is derived to match the $\coo(2-1)$ radial emission between $40$ and $300\,{\rm AU}$, and find best-fit values of the parameters $\gamma \sim 0.4$ and $r_{c} \sim 300\,{\rm AU}$. 
Then, we initially assume an arbitrary constant $\varepsilon$ to obtain the original dust surface density $\sigmadust^{\rm O}(r) = \varepsilon \ \sigmagas(r)$, $\varepsilon$ is later determined by fitting the continuum data.  
Finally, we apply the dust fragmentation and drift as described in Appendix~\ref{appendix:model_C_theories} to obtain the final dust distribution $\sigmadust(r, a)$.
We solve for fragmentation equilibrium first and then consider drift.

There are a few parameters that affect the output of the synthetic continuum images, including turbulence ($\alpha_{\rm v}$), radial drift radius ($r_{\rm drift}$), size ($a_{\rm drift}$), and fragmentation velocity ($u_{\rm frag}$).
Some of these parameters can be constrained, thus reducing the degrees of freedom.

\paragraph{$\alpha_{\rm v}$}
The turbulence parameter describes the efficiency of the dust collisional equilibrium and also affects dust settling (see Section~\ref{subsec:data_driven_model_B} for more details).
It is constrained from the infrared SED at $\sim 100\,\um$, and we use the best-fit $\alpha_{\rm v} \sim 10^{-3}$ from the structured model. 

\paragraph{$r_{\rm drift}$ and $a_{\rm drift}$}
The outer edge of the dust disk shown in the millimeter wavelength continuum images indicates how far the grains have drifted, so we set $r_{\rm drift}$ to $300\,{\rm AU}$ ($\sim R_{\rm dust}$), and find $a_{\rm drift} \sim 0.01\,{\rm cm}$ by solving Equation~\ref{equation:amax_drift}.

\paragraph{$u_{\rm frag}$}
The fragmentation velocity is expected to be a value between $100-1000\,{\rm cm\,s^{-1}}$ \citep[][]{Gundlach2018}, and a higher $u_{\rm frag}$ value will result in fragmentation of the larger grains, leaving smaller grains in the outer disk.
We find that smaller values of $u_{\rm frag} < 750\,{\rm cm\,s^{-1}}$ would result in a $\sigmadust(r,a)$ for large grains that are inconsistent with the data (see Figure~\ref{fig:Result_Model_C_dustfrag} in Section~\ref{subsec:drift_discussions} for more details), and we therefore adopt $u_{\rm frag} = 750\,{\rm cm\,s^{-1}}$ to reduce the mass in small grains that remain beyond $r_{\rm drift}$.

\begin{figure}[h]
\gridline{\fig{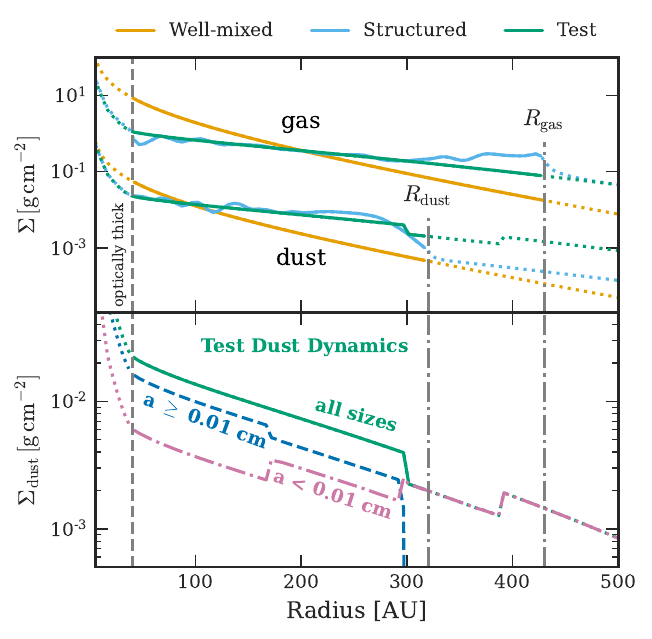}{0.47\textwidth}{}}
\caption{
The gas surface densities for the well-mixed (orange), structured (blue) and the dust dynamics test scenario (green) are shown in the top panel. 
The dotted lines represent the regions where we lack observational information to derive robust results, including the inner disk where it becomes optically thick (vertical dashed line), and the outer disk that is beyond measured dust or gas radii ($R_{\rm dust}$ and $R_{\rm gas}$ shown in vertical dash-dotted lines).
In the bottom panel, the three lines show $\sigmadust(r)$ for different grain sizes when computed using fragmentation and drift:
green line for all grain sizes, blue dashed line for larger grains ($a \geq a_{\rm drift}$) that have drifted in, and magenta dash-dotted line for smaller grains ($a < a_{\rm drift}$). There are some visible steps in the dust surface densities 
in both panels, and they are due to the effect of the dust fragmentation solved from a finite number of dust size bins.
}
\label{fig:Model_surface_density}
\end{figure}

We present the resulting surface densities for both dust and gas in the top panel of Figure~\ref{fig:Model_surface_density}.
By construction, the gas surface density is similar to that of the structured model, apart from the fact that we forced a match to the self-similar profile.
The lower panel shows $\sigmadust(r)$, where the $\sigmadust(r,a)$ for smaller ($a < a_{\rm drift}$) and larger ($a \geq a_{\rm drift}$) particles are shown as magenta dash-dotted, and blue dashed lines, respectively.
Large dust particles ($a \geq a_{\rm drift} = 0.01\,{\rm cm}$) have to drift inward of $R_{\rm dust}$ to reproduce the continuum profile, while smaller particles are coupled to the gas dynamically and remain in the outer disk.

\subsubsection{Results for the test of dust dynamical effects}

\label{subsection:model_C_result}

With the self-similar fit to $\sigmagas(r)$ and $\sigmadust(r, a)$, we run \texttt{DiskMINT} with the same setups as the structured model. 
The shape of $\sigmadust(r,a)$ is determined by solving fragmentation and drift, and we vary the total $\mdust$ to match the mm-wavelength flux density profile.
We summarize all the adopted parameters for this test in Table~\ref{Tab:model_C_para}.

\begin{figure}[h]
\gridline{\fig{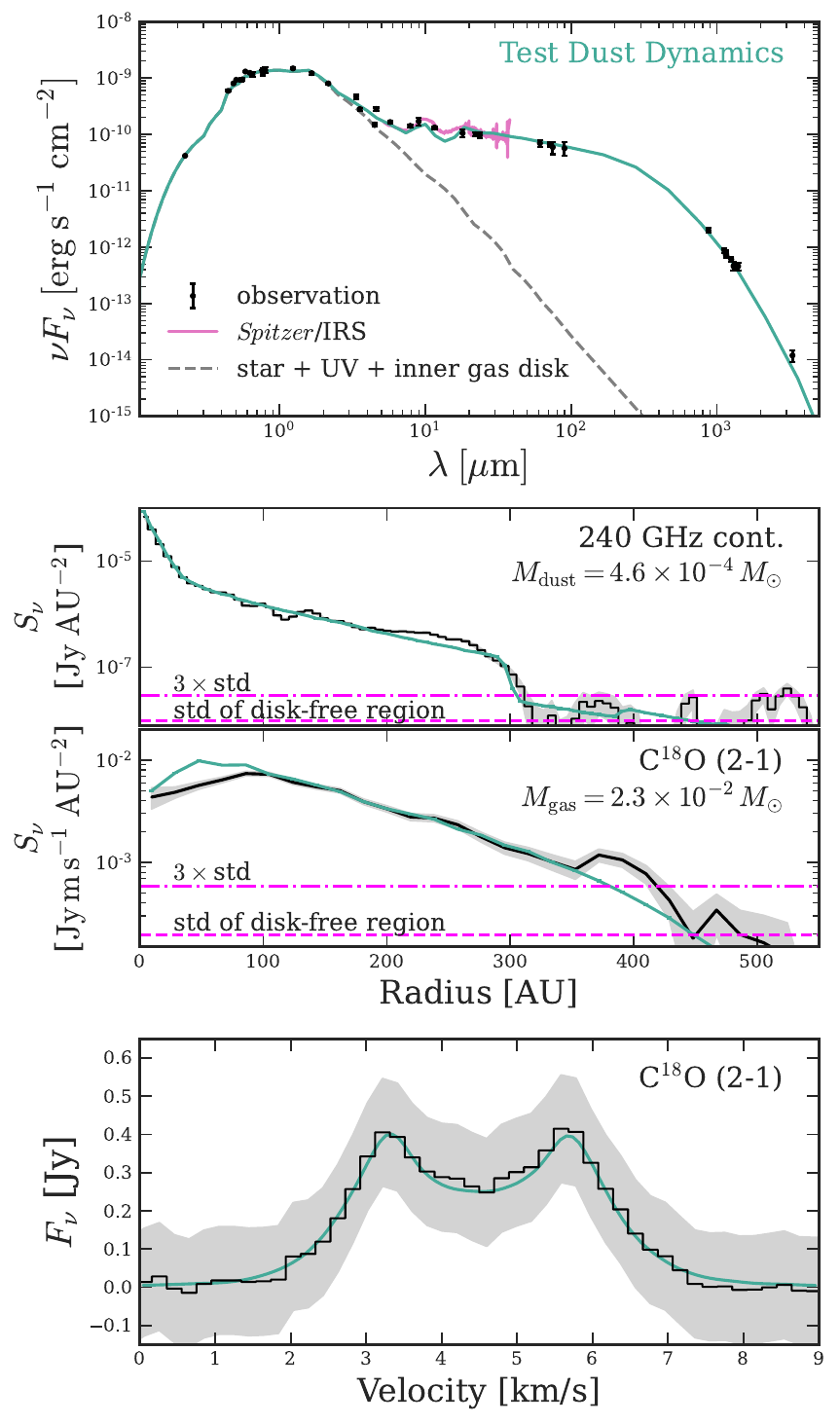}{0.45\textwidth}{}}
\caption{Compare SED (top), dust continuum and $\coo(2-1)$ radial profile (middle), and $\coo(2-1)$ line spectra between the dust dynamics test model and observation. All the notations are the same as Figure~\ref{fig:Result_Model_B}, with the green lines representing the model result and the black line showing the observation with uncertainties shaded by the gray area.}
\label{fig:Result_Model_C}
\end{figure}

We find the best-fit is $\mdust = 4.6\times 10^{-4}\,M_\odot = 153\,M_\oplus$.
We summarize the main results, including the SED fitting, and fitting of the continuum and $\coo(2-1)$ radial profiles, and the $\coo(2-1)$ line spectra in Figure~\ref{fig:Result_Model_C}.
Since $\sigmagas(r)$ is set based on that from the structured model in $40\,\mathrm{AU} \lesssim r \lesssim 300\,\mathrm{AU} (\sim R_{\rm dust})$ (solid green line), and we obtain a similar estimate of $\mgas \sim 2.3\times10^{-2}\,M_\odot$ in that radius range.
The lower total mass for this case is because we neglected structures by using a self-similar fit, and especially a mass enhancement at the outer disk $r > 300\,{\rm AU}$ (see Fig.~\ref{fig:Model_surface_density}) that is needed to match the data. 

\begin{deluxetable}{lcccc}
\tablecaption{Parameters for the Dust Dynamics Test Case
\label{Tab:model_C_para}}
\tabletypesize{\normalem}     
\setlength{\tabcolsep}{0.5pt}       
\tablewidth{0.45\textwidth}
\tablehead{
\colhead{Parameter} & \multicolumn{2}{c}{Symbol} & \multicolumn{2}{c}{Value}
}
\startdata
    \textit{Dust Properties} &&&& \\
    Composition &&& \multicolumn{2}{c}{DIANA standard dust} \\
    minimum size & \multicolumn{2}{c}{$a_\mathrm{min}$}  &  \multicolumn{2}{c}{$1\times10^{-5}$ cm} \\
    maximum size\tablenotemark{a} & \multicolumn{2}{c}{$a_\mathrm{max}$}  &  \multicolumn{2}{c}{$0.06\,\mathrm{cm}$} \\
    exponential slope\tablenotemark{a} & \multicolumn{2}{c}{$q$}  &  \multicolumn{2}{c}{3.2} \\ 
    \hline
    \textit{Vertical Structure} &&&& \\
    vertical density distribution &&& \multicolumn{2}{c}{\makecell[c]{solve VHSE with dust \\ settling in iterations}} \\
    $-$ turbulence\tablenotemark{a} & \multicolumn{2}{c}{$\alpha_{\rm v}$} & \multicolumn{2}{c}{$1\times10^{-3}$}\\
    \hline
    \textit{Radial Structure} &&&& \\
    inner radius of the disk & \multicolumn{2}{c}{$r_\mathrm{in}$}  &  \multicolumn{2}{c}{0.055 AU}\\
    gas surface density & \multicolumn{2}{c}{$\sigmagas(r)$} & \multicolumn{2}{c}{\makecell[c]{adopt Equation~\ref{eq:power-law-sigma} \\ and fit to $\sigmagas(r)$ \\ of structured model}} \\
    $-$ characteristic radius & \multicolumn{2}{c}{$r_c$}  &  \multicolumn{2}{c}{300 AU} \\
    $-$ surface density slope & \multicolumn{2}{c}{$\gamma$} & \multicolumn{2}{c}{0.4} \\
    dust surface density & \multicolumn{2}{c}{$\sigmadust(r)$} & \multicolumn{2}{c}{\makecell[c]{solve dust fragmentation \\ and drift}} \\
    $-$ radial drift radius & \multicolumn{2}{c}{$r_{\rm drift}$} & \multicolumn{2}{c}{$300\,{\rm AU}$}\\
    $-$ radial drift size & \multicolumn{2}{c}{$a_{\rm drift}$} & \multicolumn{2}{c}{$0.01\,{\rm cm}$}\\
    $-$ fragmentation velocity & \multicolumn{2}{c}{$u_{\rm frag}$} & \multicolumn{2}{c}{$750\,{\rm cm\,s^{-1}}$}\\
\enddata
\tablecomments{This is a test to evaluate dust dynamics using the gas distribution from the structured model. All parameters are the same as Table~\ref{Tab:model_B_para} except for $\sigmagas(r)$ and $\sigmadust(r,a)$, including $r_c$, $\gamma$, $r_{\rm drift}$, $a_{\rm drift}$, $u_{\rm frag}$.}
\end{deluxetable}

The dust continuum emission from our test model can provide a good match to the data (Figure~\ref{fig:Result_Model_C} second panel from top), suggesting that the observed dust outer edge could be explained by dust evolution in the disk around IM~Lup. 
The differences between the $\mdust$ estimates compared to the structured model are mostly due to the remaining dust particles beyond $R_{\rm dust}$, since we have assumed that the outer disk is described by a self-similar profile extending beyond the measured disk radii. However,
the $\varepsilon(r)$ within $r < R_{\rm dust}$ remains similar (Figure~\ref{fig:Model_surface_density}).
We note that with the smoother slope for both $\sigmagas(r)$ and $\sigmadust(r)$ compared to those in the structured model, the low-level structures are ignored (Figure~\ref{fig:Model_surface_density} top panel). 

\section{Discussions and implications} \label{sec:discussions}

In this work, we present two different disk models: the first well-mixed model reproduces the SED and the integrated $\coo$ luminosity, while the second structured model also matches the continuum emission profile and the $\coo$ radial profile.
Table~\ref{Tab:model_results} summarizes the main results and demonstrates that the models provide consistent $\mdust$ and $\mgas$ within a factor of a few.
Here, we first discuss potential uncertainties on $\mgas$ by comparing \texttt{DiskMINT} with detailed thermochemical models (Section~\ref{subsec:uncertainty_gas_mass}). 
Next, we compare our $\mgas$ estimates to values reported in the literature and further argue that $\coo$ can be used as a good $\mgas$ tracer (Section~\ref{subsec:discuss_gas_disk_mass}).
We also discuss uncertainties in $\mdust$, which is important for inferences on the gas-to-dust mass ratio (Section~\ref{subsec:discuss_dust_mass}).
Finally, we discuss the implications of our results for disk evolution and planet formation (Sections~\ref{subsec:drift_discussions} and~\ref{subsec:dust-to-gas_radial}).

\begin{deluxetable*}{c|cccc|cccc}
\tablecaption{Model Results \label{Tab:model_results}}
\tablewidth{0.90\textwidth}
\tablehead{
\multicolumn{1}{c|}{Model}  &  
\multicolumn{4}{c|}{Matches} &
\multicolumn{4}{c}{Results}\\
\multicolumn{1}{c|}{}  &  
\colhead{SED ($\gtrsim 100\,\um$)?}  & 
\colhead{cont. profile?}  & 
\colhead{$L_{\rm C^{18}O}$?}  & 
\multicolumn{1}{c|}{${\rm C^{18}O}$ profile?}  & 
\colhead{$M_{\rm dust}$}  &
\colhead{$M_{\rm gas}$} & 
\colhead{$\varepsilon$} & 
\colhead{g/d} \\
\multicolumn{1}{c|}{}  & 
\colhead{}  & \colhead{}  & 
\colhead{}  & \multicolumn{1}{c|}{}  & 
\colhead{($M_\oplus$)}  & 
\colhead{($M_\odot$)} & 
\colhead{} &
\colhead{}
}
\startdata
Well-mixed & \cmark & -- & \cmark & -- & $83$ &  $0.039$ & $6.41\times10^{-3}$ & 156 \\
Structured & \cmark & \cmark & \cmark & \cmark &  $133$ & $0.028$ & $1.43\times10^{-2}$ & 70 \\ 
\enddata
\tablecomments{Here we present the best-fit $\mgas$ and $\mdust$ values for each model. We discuss associated uncertainties for the higher fidelity structured model using our method in Sections~\ref{subsec:discuss_gas_disk_mass} and~\ref{subsec:discuss_dust_mass}.}
\end{deluxetable*}

\subsection{Gas masses constrained within a factor of a few}
\label{subsec:uncertainty_gas_mass}

Here, we discuss two main sources of uncertainty in our gas mass estimates: i) the omission of gas thermal processes in \texttt{DiskMINT} and ii) the limited observational constraints in the optically thick inner disk and the faint outer disk. While both contribute to underestimating $\mgas$, we show the effect is limited to within a factor of $\sim 2$.

\paragraph{Omission of gas cooling}
This work uses the dust-based model \texttt{DiskMINT} \citep{Deng_2023_diskmint}, a simplified version of the thermochemical model from \citet{Ruaud2019} and \citet{Ruaud2022}. 
Specifically, gas-phase heating and cooling mechanisms are omitted and $T_{\rm gas}$ is computed from the thermal equilibrium with dust grains of different sizes ($T_{\rm gas} \sim \left<T_{\rm dust}(a)\right>$).
This simplification saves a significant amount of time -- One \texttt{DiskMINT} model requires $\sim 60$ CPU (2.1 GHz) Hours, while the full model takes at least 10 times as much.
As demonstrated in \citet{Ruaud2022}, this simplified approach yields a thermal profile relatively similar to the full model, particularly near the disk midplane, where the dust density is high and where the $\coo$ emission arises (see Figure~9 in \citealt{Ruaud2022}).
The largest differences are for the lower-mass disks ($\mgas \lesssim 1\times10^{-3}\,M_{\odot}$), where low densities in the $\coo$ emitting layer can make gas cooling significant. 
In these disks, the simplified model can overestimate the $\coo$ line fluxes (hence underestimate the mass) by up to a factor of $\sim 2$ compared to the full model in \citet{Ruaud2022}. 
However, the disk of IM~Lup is relatively massive, and densities are high enough that gas heating and cooling are dominated by dust collisions.
Therefore, a detailed thermo-chemical model is expected to yield a mass at most larger than a factor of $\sim2$ compared to \texttt{DiskMINT}, and an approximate upper limit to IM~Lup $\mgas$ can be set as $\sim 2\times\mgas$, corresponding to $\sim 0.08\,M_\odot$.

\paragraph{The optically thick inner disk and faint outer disk}

In this work we have expanded upon the publicly available version of \texttt{DiskMINT} by enabling fits to both dust and gas radial profiles. 
However, it is important to keep in mind that even tracers typically considered optically thin may not fully probe the disk, as optical depth increases closer to the star and sensitivity drops at larger radii.

In the specific case of IM~Lup, the dust and gas become optically thick inside $\lesssim 40\,{\rm AU}$, hence what we derive within is a lower limit to the surface density.
For example, in the structured model, the integrated mass in this optically thick region is $\sim 2.7\times10^{-3}\,M_{\odot}$, $< 10\%$ of the total gas mass. 
We therefore conclude that this region will not significantly contribute to the total gas mass derived in the structured model.
Additionally, because this region is optically thick, our estimates of the $\mgas$ could potentially be a lower limit inside $\lesssim 40\,{\rm AU}$.

In the outer disk, sensitivity sets a limit at $\sim 300\,{\rm AU}$ for dust and at $\sim 400\,{\rm AU}$ for $\coo$ gas, thus we have no direct observational constraints at larger radii. 
However, the continuum emission declines sharply beyond $\sim 300\,{\rm AU}$, indicating that the data (and hence our modeling) capture most of the outer dust mass. 
Since our gas temperature is based on dust, there needs to be some small dust in the outer region as well, and the structured model includes a small component beyond the continuum edge that does not produce a detectable $F_{\rm mm}$. 
This sets the gas temperature in the region from $\sim 300-400\,{\rm AU}$ and the gas mass and surface density here is constrained by the $\coo$ radial profile. 
The gas emission traced by $\coo$, however, shows no obvious drop at $\sim 400\,{\rm AU}$, and the main isotope extends out to much larger radii. 
As such, the structured model grid extends out to $1,000\,{\rm AU}$. 
The mass from $\sim400-1000\,{\rm AU}$ (which consists of approximately 25\% of the total $\mgas$) is therefore a little uncertain. 
We find that fitting the $\coo$ radial emission in the structured model necessitates a region of enhanced surface density near $\sim 400\,\mathrm{AU}$, which could have a physical explanation in the presence of spiral arms in the outer disk \citep[][]{huang_disk_2018}.
The outer disk beyond $\sim 400\,\mathrm{AU}$ could be traced by $\twelveco$ and $\thirteenco$, and they could potentially be used to constrain $\sigmagas(r)$ when $\coo$ emission is not detected.
However, both $\twelveco$ and $\thirteenco$ lines are still optically thick at $\sim 400\,{\rm AU}$ and only become optically thin at even larger radii where $\sigmagas(r) \lesssim 0.1\,{\rm g\,cm^{-2}}$ for $\thirteenco$, and $\sigmagas(r) \lesssim 0.05\,{\rm g\,cm^{-2}}$ for $\twelveco$. In addition, both lines emit at higher layers ($z/r \gtrsim 0.3$) where gas cooling cannot be omitted and is associated with kinematic structures \citep[e.g.,][]{huang_disk_2018}.
As such, more detailed models are needed to properly infer $\sigmagas(r)$ from those two tracers in the outer disk, which could be explored in future work.
Based on modeling the $\coo$ line emitting region alone, we conclude here that the IM~Lup $\mgas$ cannot be lower than $\sim 0.02\,M_\odot$, which corresponds to $\sim 75\%$ of the total $\mgas$.

\subsection{\texorpdfstring{$\coo$: a reliable $\mgas$ tracer with self-consistent models}{C18O: a reliable Mgas tracer with self-consistent models}}
\label{subsec:discuss_gas_disk_mass}

Several methods have been proposed in the literature to estimate gas disk masses \citep[e.g.,][for a recent review]{Miotello_PPVII_2023}. 
The massive and well-resolved disk of IM~Lup has enabled exploration of all of them. 
As a result, IM~Lup has $\mgas$ estimates derived from three distinct approaches:
\begin{enumerate}[label=\alph*)]
    \item Dust-based estimate: the simplest among all methods is to obtain $\mgas$ by multiplying the dust mass ($\mdust$) by a constant gas-to-dust mass ratio, typically 100 as in the ISM (\citealt{pinte_probing_2008, cleeves_IMLup_coupled_2016, zhang_MAPS_molecules_2021}).
    \item Dynamical mass estimate: by modeling the high-resolution $\twelveco$ (and/or $\thirteenco$) rotation curves \citep{lodato_dynamical_mass_2023, martire_rotation_2024}. This is considered to be the most reliable method for massive disks, but it requires assuming a disk temperature structure.
    \item Thermochemical modeling: there are two variations. The standard method is by modeling the line intensities (and also radial profiles in some works, including this one) of less abundant CO isotopologues \citep[][and this work]{zhang_MAPS_molecules_2021, trapman_exoalma_2025}. A more recent method models the vertical emitting height (and gas temperature) of the optically thick $\twelveco$ (and/or $\thirteenco$) line many scale heights above the disk to make inferences on the total $\mgas$  \citep[][]{paneque-carreno_vertical_2025, rosotti_exoalma_2025}.
\end{enumerate}

\begin{figure*}[ht]
\gridline{\fig{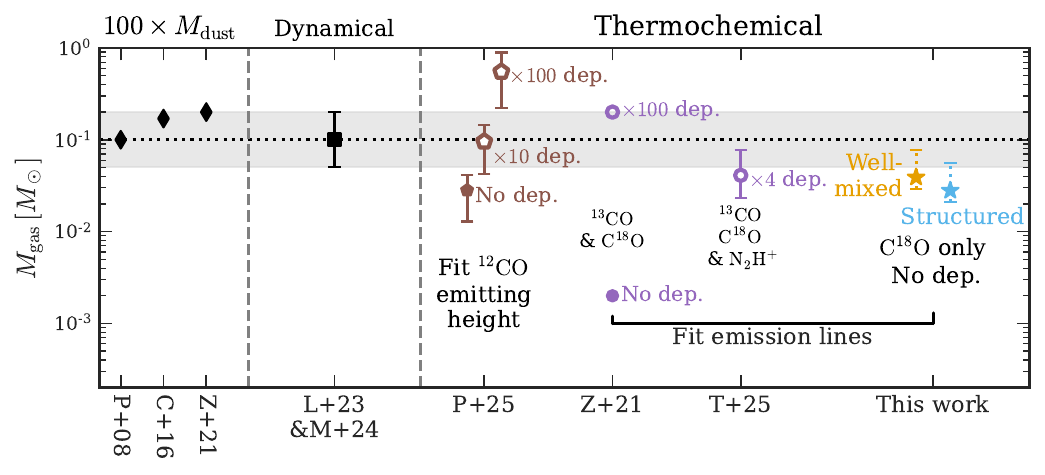}{0.95\textwidth}{}}
\caption{$\mgas$ estimates in the literature.
From left to right, they include $\mgas$ based on $\mdust$ multiplied by a constant ISM gas-to-dust mass ratio of 100 from P+08\citep[][]{pinte_probing_2008}, C+16\citep[][]{cleeves_IMLup_coupled_2016}, Z+21\citep[][]{zhang_MAPS_molecules_2021}, the dynamical $\mgas$ by L+23 \& M+24 \citep[][]{lodato_dynamical_mass_2023, martire_rotation_2024}, the $\mgas$ inferred from the $\twelveco$ emitting height by P+25\citep[][]{paneque-carreno_vertical_2025}, and the $\mgas$ inferred from the CO isotopologue line fluxes by Z+21\citep[][]{zhang_MAPS_molecules_2021}, T+25\citep[][]{trapman_exoalma_2025}, and this work.
The solid pentagons and circles represent models with no assumed CO or C depletion, and the hollow ones are derived with a depletion factor that is annotated alongside.
The dotted black horizontal line shows the median value of the derived dynamical mass, with the shaded gray region for its uncertainty.
$\mgas$ is better constrained by recent works: \citet{trapman_exoalma_2025} used a combination of $\thirteenco$, $\coo$ \& ${\rm N_2H^{+}}$ line emission, and this work utilizes $\coo$ line emission only. 
Both works find $\mgas$ estimates that are consistent with the inferred dynamical $\mgas$ within quoted uncertainties.
For the estimates obtained in this work, the lower and upper limits are marked as $\times0.75$ and $\times2$, respectively, and are shown as dashed error bars because they represent the uncertainties associated with the method, rather than the statistical errors quoted for the points in the literature.
}
\label{fig:Mgas_comp_literature}
\end{figure*}

Figure~\ref{fig:Mgas_comp_literature} compares results from these three methods on the IM~Lup disk.
The dust-based $\mgas$ (black diamonds) and the dynamical masses (black squares), shown in the first two panels, are consistent within the reported uncertainties. This suggests that the gas-to-dust mass ratio in IM~Lup is close to the canonical ISM value, in agreement with other large Myr-old disks \citep{pascucci_noCOdep_2023}. 
The estimates of $\mgas$ derived from thermochemical disk models are presented in the last panel.
The solid pentagons and circles are without considering any homogeneous CO or C depletion factors, while the hollow ones are literature values assuming different depletion factors so that the model gas disk masses align with dust-based estimates \citep[][]{zhang_MAPS_molecules_2021}, dynamical estimates \citep[][]{paneque-carreno_vertical_2025}, or additional emission lines (e.g., ${\rm N_2H^+}$, \citealt{trapman_exoalma_2025}).
A temporal trend is evident, with more recent CO-based estimates yielding higher gas masses for IM~Lup, implying less CO depletion. 
Our $\mgas$ estimates, obtained with an ISM CO abundance as input, are consistent with the dynamical mass estimate, particularly when considering the likely factor of two underestimation due to the omission of gas cooling in \texttt{DiskMINT} (see previous Subsection~\ref{subsec:discuss_gas_disk_mass}). 

How can we explain the apparent evolution in $\mgas$ and CO depletion estimates from thermochemical models?
The earliest and lowest $\mgas$ value in Figure~\ref{fig:Mgas_comp_literature} is from \citet{zhang_MAPS_molecules_2021} using \texttt{RAC2D} \citep[][]{du_water_2014}. 
They did not consider any of the three processes we identify as crucial to determining gas masses from $\coo$ \citep[][]{Ruaud2022}; isotope-selective dissociation, freeze-out with grain-surface chemistry, or vertical hydrostatic equilibrium with consistent density-temperature structure.
They further did not recompute the chemistry after imposing a global CO depletion (see the discussion in Section~3.3 of \citealt{pascucci_noCOdep_2023}). 
Therefore, they need a high depletion factor to reconcile their model masses. 
\citet{paneque-carreno_vertical_2025} and \citet{trapman_exoalma_2025} both used \texttt{DALI} \citep[][]{bruderer_warm_2012, bruderer_survival_2013}, which includes isotope-selective dissociation \citep[][]{Miotello2014, Miotello2016, miotello_lupus_2017} and a grid of pre-computed disk models with varying degrees of global CO depletion (hence the purely statistical uncertainties shown in Figure~\ref{fig:Mgas_comp_literature}). 
\citet{paneque-carreno_vertical_2025} re-discover the need of vertical hydrostatic equilibrium; recognized as essential physics by earlier disk models \citep[e.g.,][]{dalessio_accretion_1998, kamp_gas_2004, gorti_line_2008} but abandoned in many recent disk models. 
We note that we previously emphasized the need for a consistent density structure in order to properly estimate $\mgas$ \citep[][]{Ruaud2022}, and this is included in \texttt{DiskMINT} \citep{Deng_2023_diskmint}. 
However, \citet{paneque-carreno_vertical_2025} used the VHSE \texttt{DALI} models to match the emitting layers of $\twelveco$ and $\thirteenco$ lines, which arise high in the disk surface near the photodissociation front or photodissociation region (PDR) layer. 
They inferred a factor of 10 depletion in CO to match the dynamical mass of IM~Lup, but this method relies on optically thick lines and is sensitive to disk temperature, which is now coupled to the density. 
In contrast, \citet{trapman_exoalma_2025} did not assume vertical hydrostatic equilibrium in their model grid, but instead used two optically thin tracers, $\coo$ and ${\rm N_2H^+}$. 
The \texttt{DALI} model does not include grain-surface chemistry -- conversion of CO to CO$_2$ ice as the main reaction -- which raises the vertical snowline. 
As the density drops sharply with height in a disk, they homogeneously deplete the CO in the disk, but only by a factor of 4 for IM~Lup, to be able to also match the ${\rm N_2H^+}$ line, which originates below the CO freeze-out layer. 

In our self-consistent \texttt{DiskMINT} models, the only input variables are the gas and dust surface density distributions; the CO abundance is not treated as a free parameter and is computed from the chemical network. 
As shown in Figure~\ref{fig:Mgas_comp_literature}, these models find gas masses that closely match the dynamical mass, without requiring adjustments to the input CO abundance or introductions of global depletion parameters. 
We note that dynamical mass estimates rely on assumptions about temperature profiles extrapolated to the disk midplane. 
For example, \citet{martire_rotation_2024} showed the sensitivity of the dynamical mass estimate to temperature: a higher surface temperature at the CO emitting layer would result in a lower $\mgas$ estimate.
However, their assumed $T_{\rm gas}$ of the CO emitting layer ($\sim 35\,{\rm K}$ at $100\,{\rm AU}$) is indeed colder than what we found in our self-consistent thermochemical disk models; therefore, their reported median $\mgas$ estimate might be overestimated, and the true $\mgas$ could be lower.
On the other hand, the best-fit $\mgas$ estimates in our work could also be underestimated due to the reasons we discussed in Section~\ref{subsec:discuss_gas_disk_mass}, and also there could be room for very small degrees of CO depletion -- if it exists, within a factor of $\lesssim 2$ \citep[][]{Ruaud2022, pascucci_noCOdep_2023}.
Nonetheless, the $\mgas$ estimates derived from different methods converge within a factor of $\lesssim 2-3$, which is within the typical level of uncertainty and suggests that the estimates are robust to this degree of accuracy.

We conclude that our results strongly support the use of $\coo$ alone as a reliable tracer of gas mass when used within a self-consistent modeling framework.
Isotope-selective photodissociation, one of the three model ingredients we advocate, is well known as being important \citep[][]{visser_photodissociation_2009, Miotello2014, Miotello2016, miotello_lupus_2017}. 
With new high-resolution data from ALMA providing constraints on the CO emitting layers, the need for VHSE to obtain accurate density structures is being recognized. 
JWST observations of ices in edge-on disks are now revealing the presence of mixed CO/CO$_2$ ice mantles on water ice-coated grains \citep[][]{sturm_edge-protoplanetary_2023, bergner_jwst_2024}, providing support for the grain-surface chemistry -- CO/CO$_2$ ice conversion as the main reaction -- we advocate in determining the CO snowline. 
We also present the model results for the ALMA Band~3 continuum and $\coo\,(1-0)$ line in Appendix~\ref{appendix:model_B_band3}. 
This line traces gas even closer to the disk midplane compared to the $\coo\,(2-1)$ line. The good match to this longer wavelength data supports our computation of the location of the CO snowline.
Finally, a similar conclusion on $\coo$ as an excellent mass tracer has been reached recently by \citet{zwicky_dancing_2025} with their independent self-consistent thermochemical disk models that have all three necessary processes as included here. In addition, a similar conclusion is reached by comparing gas masses inferred from $\thirteenco$, $\coo$ and ${\rm N_2H^+}$ \citep[][]{trapman_exoalma_2025, Trapman_AGEPRO_V_gas_masses} with the grid of self-consistent models from \citet{Ruaud2022}; see Figure~14 in \citet{Deng_2025_AGEPRO_III_Lupus}.

\subsection{Uncertainties on the dust disk mass}
\label{subsec:discuss_dust_mass}

Most dust disk mass estimates in the literature are obtained from one millimeter continuum flux measurement under the assumption that the emission is optically thin:

\begin{equation}
    \mdust = \frac{B_\nu(\bar{T}_{\rm dust})\kappa_\nu}{{d^2}} F_{\rm cont. \nu},
\end{equation}
where $\bar{T}_{\rm dust}$ is the average dust temperature, $\kappa_\nu$ is the dust opacity, and $d$ is the distance from the observer to the disk. 
Obtaining analytical $\mdust$ estimates requires assumptions on both  $\bar{T}_{\rm dust}$ and $\kappa_\nu$.
Although obviously approximate, this simple method is capable of providing first-order $\mdust$ estimates and has been utilized in many previous works \citep[e.g.,][]{Hildebrand_dustmass_1983, Andrews_n_Williams_dustmass_2005, Hendler2017HintsDwarfs}.
Typical adopted parameters are $\bar{T}_{\rm dust} = 20\,K$ for the average dust temperature and $\kappa_\nu = 2.3\,{\rm cm^2\,g^{-1}}$ for observations at at 230\,GHz \citep[e.g.,][]{Ansdell_lupus_2016, Pascucci_mass_2016}. 
With these assumptions, the dust disk mass of IM~Lup is $\mdust \sim 137\,M_\oplus$ \citep[][]{Manara2022DemographicsFormation} with the revised Gaia distance.

Modeling the full SED, as done in this work, enables a self-consistent determination of the dust temperature and optical depth, resulting in a more reliable estimate of the dust disk mass \citep[see also][]{ballering_protoplanetary_2019}. 
Using this approach, \citet{zhang_MAPS_molecules_2021} found $\mdust = 1.97\times10^{-3}\,M_\odot = 656\,M_{\oplus}$, a factor of $\sim 5$ higher than that reported in the literature. 

In this work, we find the following range $\mdust \sim 83 - 153\,M_\oplus$, in agreement with the optically thin approach. 
Indeed, our detailed modeling finds an optical depth $\tau \lesssim 0.01$ beyond $r \gtrsim 50\,{\rm AU}$.
The difference in mass is mainly due to the difference in the assumed opacities; \citet{zhang_MAPS_molecules_2021} used DSHARP values \citep[][]{birnstiel_disk_2018} which are a factor of $\sim 5$ lower at mm-wavelengths than the DIANA standard values used in this work and in \citet{Manara2022DemographicsFormation}. 
In a more detailed analysis, \citet{sierra_molecules_2021} inferred $\mdust$ and $\sigmadust(r)$ from dust evolutionary models fitting ALMA continuum observations at four different frequencies at both Bands~3 and~6.
They also adopted the DSHARP dust composition and inferred $\mdust \sim 639\,M_{\oplus}$, which is similar to the value derived in \citet{zhang_MAPS_molecules_2021} using the same dust composition.
Interestingly, they also inferred a relatively flat $\sigmadust(r)$ beyond $r \gtrsim 50\,{\rm AU}$ (Figures~7\&11 in \citealt{sierra_molecules_2021}), which agrees with what we find here -- except their absolute value is a factor of $\sim5$ higher compared to this work, mainly due to the factor of $\sim 5$ difference in adopted opacities. 
If the dust in the IM~Lup disk has been processed to the extent that it resembles meteoritic compositions, which the DSHARP values are based on, then our dust mass estimates could be underestimated by a factor of $\sim 5$. 
We note that the small dust component has to be similar to the ISM composition (DIANA opacities) to match many dust emission features.
Indeed, many models include two populations with differing compositions, as do \citet{zhang_MAPS_molecules_2021}. 
If the dust is in collisional/fragmentation equilibrium, it is not clear how these two distinct populations can be maintained. 
However, a discussion of dust compositions and their evolution in disks is beyond the scope of this work. 
In our simplified modeling, we also ignore the effects of ice mantles in the midplane on dust opacities in the radiative transfer, although these effects are included in the chemistry. 
This would lower the opacities and increase our dust mass estimates by $\sim 20$\% as well. 
These effects will be explored in future work.

This comparison between $\mdust$ estimates in this work with the ones reported in the literature indicates that uncertainties in dust properties that determine the opacity are the primary source of uncertainty. 


\subsection{Indications of radial drift in the IM~Lup disk}
\label{subsec:drift_discussions}
As commonly observed in most disks, the radial extent of the IM~Lup dust disk as probed by the sub-millimeter continuum is significantly smaller than that of the gas disk and the NIR dust disk.  
Since the emission at a given wavelength is dominated by particles of a similar size ($ 2\pi a \sim \lambda$), large dust particles are absent beyond the dust outer edge $R_{\rm dust}$. 
This could be due to fragmentation and drift \citep[e.g.,][]{birnstiel_simple_2012}, which were both considered in our test of dust dynamics (see \ref{subsection:model_C_result}). 

\begin{figure}[ht]
\gridline{\fig{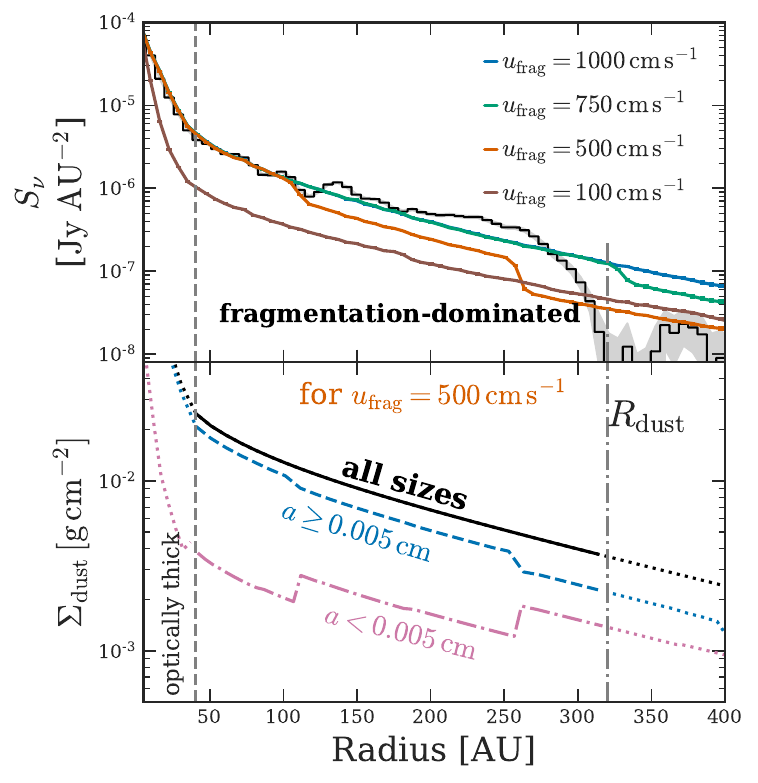}{0.45\textwidth}{}}
\caption{The dust continuum radial distribution with different $u_{\rm frag}$ (top) and $\sigmadust(r)$ (bottom) for the model with $u_{\rm frag} = 500\,{\rm cm\,s^{-1}}$. The models presented here are with fragmentation-dominated grain size distributions, and we do not include the dust drifting. In the top panel, we plot the observed continuum radial distribution in black with its uncertainties in gray shade, and the models with $u_{\rm frag}=1000, 750, 500, 100\,{\rm cm\,s^{-1}}$ are shown in blue, green, orange, and brown, respectively. In the bottom panel, the $\sigmadust(r)$ for all dust grains, large grains ($a \geq 0.005\,{\rm cm}$, which is responsible for the millimeter-wavelength continuum emission), and small grains ($a < 0.005\,{\rm cm}$) are shown as black solid, blue dashed, and pink dash-dotted lines, respectively. The gray vertical dashed line marks the region where the dust becomes optically thick, and the gray vertical dashed-dotted line shows the dust outer radius $R_{\rm dust}$.}
\label{fig:Result_Model_C_dustfrag}
\end{figure}

Fragmentation alone can introduce an mm-flux edge that differs from that of the gas. 
The maximum grain size $a_{\rm max}$ decreases with radius as $\sigmagas$ decreases (see Equation~\ref{equation:fragmentation_size} in Appendix~\ref{appendix:model_C_theories}). 
Therefore, it is to be expected that at some outer radius, grains do not grow to sizes needed for strong millimeter emission, and the mm-flux drops while gas and small dust particles remain. 
From our structured model, we estimate $\sigmagas/c_s^2$ and vary $u_{\rm frag}$ and test the fragmentation-only scenario. 
We show the resulting continuum radial distribution of this test model for four different values of $u_{\rm frag}=100, 500, 750\,{\rm and}\,1000\,{\rm cm\,s^{-1}}$ in the top panel of Figure~\ref{fig:Result_Model_C_dustfrag}. 
For clarity, we pick one of these ($u_{\rm frag}=500\,{\rm cm\,s^{-1}}$) to show the model dust surface densities in the bottom panel. 
For all three values, the mm-emission in the disk beyond $R_{\rm dust}$ is not suppressed to the extent indicated by the ALMA data. 
In this model without radial drift, the decrease of the millimeter-wavelength continuum radial profile is also shallower and cannot reproduce the observed sharp drop at $R_{\rm dust}$. 
We therefore conclude that the difference in dust and gas disk radii of the IM~Lup disk is not easily explained with a dust evolution model that includes only fragmentation. 

\begin{figure}[ht]
\gridline{\fig{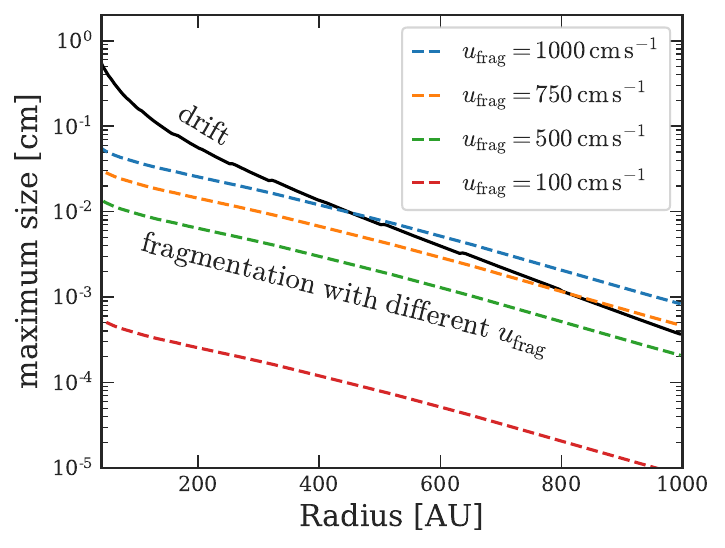}{0.45\textwidth}{}}
\caption{The maximum size ($a_{\rm max}$) that the dust grain could live as limited by the drift (black solid line) or fragmentation (color dashed lines). $a_{\rm max}$ at different radii are constrained by the smaller value solved from either drift or fragmentation (with a certain $u_{\rm frag}$), and they are solved based on the surface density and the thermal profile for the validation model. The fragmentation is always the dominant size-constraining factor in the inner disk and could also dominate in the outer disk if $u_{\rm frag} < 750\,{\rm cm\,s^{-1}}$. Drift could only become the dominant factor at the outer disk with relatively large $u_{\rm frag} \gtrsim 750\,{\rm cm\,s^{-1}}$}
\label{fig:size_threshold}
\end{figure}

We find that the mm-dust edge is well explained by considering radial drift, as shown earlier (see Section~\ref{subsec:theoretical_model_C}). 
There, the dust drift radius was fixed to be the observationally constrained radius $R_{\rm dust}$, and fits to ALMA data provided values of $\epsilon=0.02$ and $u_{\rm frag}=750\,{\rm cm\,s^{-1}}$. 
The maximum grain size is now drift-dominated (see Equation~\ref{equation:amax_drift}, and \citealt{birnstiel_simple_2012}) in the outer disk, see Fig.~\ref{fig:size_threshold}. 
In our framework, where $T_{\rm gas}$ is determined by $T_{\rm dust}$, $a_{\rm drift}$ is only weakly sensitive to gas via the pressure gradient exponent which is set by our determinations of $\Sigma_{gas}$ and $c_s$. 
$a_{\rm frag}$, on the other hand, is proportional to the gas surface density and for the IM~Lup disk, sets the maximum grain size at $r\lesssim300\,{\rm AU}$. 

We estimate the fraction of dust mass that needs to have drifted inside $r_{\rm drift}$ in our test model with dust dynamics.
We find that before applying dust radial drift, there is $\sim 100\,M_\oplus$ inside $r_{\rm drift}$, and there is $\sim 109\,{\rm M_\oplus}$ after the drift is applied, suggesting that $\sim 8\%$ of the mass ($\sim 9 M_\oplus$) that is currently inside could have drifted from outside of $r_{\rm drift}$. 
Because of this radial drift in the early stages, dust masses are slightly enhanced, resulting in a slightly larger dust-to-gas mass ratio, which may help form planetesimals.
However, the mass that is estimated to have drifted inward is too little to affect the formation of rocky planets via pebble accretion \citep[e.g.,][]{lambrechts_forming_2014}.

The extent to which drift could have occurred can be estimated as
\begin{equation}\label{equation:deltaR_drift}
    \Delta R_{\rm drift} \sim \tau_d\ v_{\rm drift},
\end{equation}
where $\tau_d \sim 1\,{\rm Myr}$ is the age of the IM~Lup system, and the drift velocity $v_{\rm drift} = -f_{\rm d} \varepsilon V_{\rm K}$.
Here $f_{\rm d}$ is an order-of-unity constant introduced in \citet{birnstiel_simple_2012} and was taken to be $0.3$ according to the calibration to more comprehensive numerical simulations \citep{birnstiel_outer_2014}, $\varepsilon$ is the dust-to-gas mass ratio, and $V_{\rm K}$ is the Keplerian velocity. 
Then, if we assume the dust-to-gas mass ratio starts from the ISM value $\varepsilon \sim 0.01$, both the gas and dust are well-coupled with the initial radius of $R^{o} \sim 1000\,{\rm AU}$, and the disk evolved for $\sim 1\,{\rm Myr}$, Equation~\ref{equation:deltaR_drift} gives $\Delta R_{dr}\sim 625\,{\rm AU}$.
This would result in an approximate dust disk outer edge at $R^{1\,{\rm Myr}} \sim R^{o} - \Delta R_{\rm drift} \sim 375\,{\rm AU}$, which is similar to $R_{\rm dust}=320\,{\rm AU}$ adopted in the modeling. 
Note that this is an order-of-magnitude consistency estimate. 
IM~Lup is observed to have spiral arms in the outer disk, while we have assumed an axisymmetric distribution, and the gas disk is unlikely to have remained static but evolved to its present configuration from when it was first formed. 
However, our radial surface density distribution determinations indicate that large dust particles have drifted inward, resulting in the observed line and continuum extent.
Detailed analysis combining dust evolution models with the thermochemical models to derive $\sigmadust(r,a)$ and $\sigmagas(r)$  is a promising future venue to obtain more accurate estimates of drift and fragmentation in disks.

\subsection{Dust-to-gas mass ratio and planetesimal formation}
\label{subsec:dust-to-gas_radial}

Streaming instability (SI) is one of the leading theories for planetesimal formation in protoplanetary disks \citep[][]{youdin_streaming_2005}, where dust particles are efficiently and rapidly collected into self-gravitating over-dense regions, thus overcoming many barriers to slow growth by collisions. 
Dust-to-gas mass ratios, often termed metallicity ($Z \sim \sigmadust/\sigmagas$), need to be high in order to trigger SI in general, but can be as low as $Z \sim 0.02$ for some particle sizes and conditions \citep[e.g.,][]{li_thresholds_2021}. 
Clumping due to SI can be more efficient when turbulence is weaker, but non-zero turbulence was found in the IM~Lup disk, and the turbulence seems to be higher at the disk's upper layers. 
For example, \citet{flaherty_evidence_2024} found $0.03 \lesssim \alpha_{\rm v} \lesssim 0.08$ by modeling the rotation curves of CO isotopologues that emit at disk surface; \citet{franceschi_IMLup_constraining_2022} found $2.1\times10^{-3} \lesssim \alpha_{\rm v} \lesssim 3.4\times10^{-3}$ by modeling the multi-wavelength observations regarding the dust vertical structure with settling; and \citet{jiang_grain-size_2024} found $\alpha_{\rm v} \sim 10^{-3}$ by modeling the grains in the pebble-settled midplane.
The latter two values are consistent with what we find in this work -- where we also mostly utilize the dust disk to constrain $\alpha_{\rm v}$ -- and this should be most relevant to the dust clumping in the disk midplane.
Therefore, the $Z \sim 0.02$ threshold inferred with $\alpha_{\rm v} \sim 10^{-3}$ \citep[][]{li_thresholds_2021} is a good approximation for the IM~Lup disk.

\begin{figure}[h]
\gridline{\fig{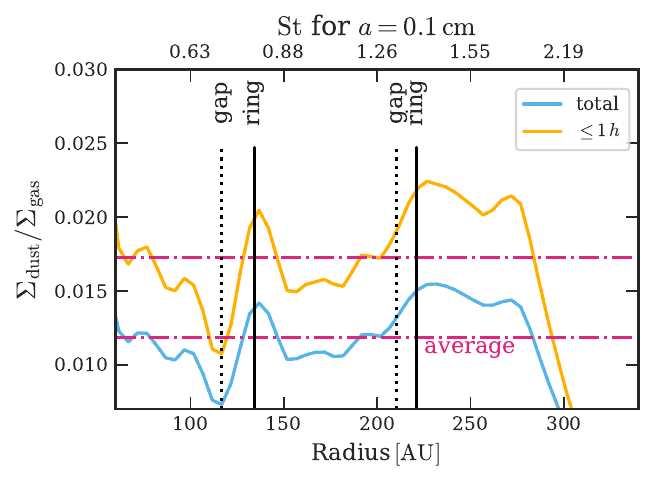}{0.45\textwidth}{}}
\caption{The dust-to-gas surface density ratio for the structured model, with the Stokes number at the disk midplane for the $0.1\,{\rm cm}$ grains shown at the top axis. 
We show the total surface densities in blue, the surface densities that are only up to one pressure scale height ($1\,h$) in orange, and the average of the two lines in red dotted-dash lines. 
We also show the dust gaps and rings identified in \citet{law_IMLup_molecules_vertical_2021} in dotted and solid vertical black lines, respectively.}
\label{fig:dust-to-gas_ratio}
\end{figure}

We plot the dust-to-gas mass ratio (metallicity) for the structured model in Figure~\ref{fig:dust-to-gas_ratio}, where we also show Stokes number in the midplane for the large grains ($\sim 0.1\,{\rm cm}$) at the top axis.
The ratio is approximately constant throughout the outer disk, even in regions where the Stokes number is relatively high. 
We note that this ratio is higher if we consider only the region of the disk within one pressure scale height $1\,h$, and it could potentially be as high as 0.1 for a different choice of dust opacity -- the derived $\sigmadust(r)$ could be a factor of $\sim 5$ higher with that smaller $\kappa_\nu$ (see Section~\ref{subsec:discuss_dust_mass} for more details) -- but the derived metallicity is still likely small enough that SI may not be triggered for a disk with a moderate level of turbulence. 
We see hints of a small enhancement in the ratio near the dust gaps and rings, but this may not be significant enough to trigger SI even in these regions. 
Therefore, our modeling results for IM~Lup do not provide support for SI condition.
The Stokes number is, though, relatively high in the outer disk for mm-sized particles, suggesting that radial drift in the outer disk is significant, and could lead to potential pile-ups as the disk continues to evolve. 

\section{Summary}
\label{sec:summary}

In this work, we present self-consistent thermochemical disk models built with \texttt{DiskMINT} (Disk Model for INdividual Targets, \citealt{Deng_2023_diskmint}), extending its initial framework to allow spatially decoupled gas and dust distributions.
We apply \texttt{DiskMINT} to IM~Lup, a large and massive disk which was reported to have an unusually low CO-based gas mass, hence inferred CO depletion by a factor of $\sim 100$ with respect to the ISM \citep[e.g.,][]{huang_disk_2018, zhang_MAPS_molecules_2021}.
Our main results are as follows:
\begin{itemize}
    \item We find $\mgas \sim 0.02-0.08\,M_\odot$ for the IM~Lup disk. Considering the underlying uncertainties, this range of values agrees with dynamical-based estimates \citep[][]{lodato_dynamical_mass_2023, martire_rotation_2024} as well as with a new thermochemical model-based estimate using a combination of CO isotopologues and ${\rm N_2H^+}$ emission \citep[][]{trapman_exoalma_2025}. 
    \item We demonstrate that the new \texttt{DiskMINT} can also derive the dust and gas surface density, hence the dust-to-gas mass ratio, from fitting the radial profiles of the dust continuum and $\coo$ line emissions. For the specific case of IM~Lup, we find a dust-to-gas mass ratio of $\sim 0.02$ between $\sim 40 -300\,{\rm AU}$ (from where dust and $\coo$ emission become optically thin to the dust disk radius) that is nearly constant with small variations within a factor of $\lesssim 2$. 
    \item We also test our derived $\sigmadust(r)$ and $\sigmagas(r)$ with dust fragmentation and drift theories using the new \texttt{DiskMINT}. For the specific case of IM~Lup, we find that the fast decline of the mm continuum emission beyond $\sim 250$\,AU is most likely shaped by drift. 
\end{itemize}

Our finding that the gas mass of IM~Lup is consistent with the dynamical mass within the described uncertainties -- without requiring global CO depletion -- supports the use of optically thin $\coo$ emission as a reliable tracer of $\mgas$ when interpreted with self-consistent models like those produced by \texttt{DiskMINT} \citep[see also][]{Ruaud2022, Deng_2023_diskmint}. 
The similarity between our disk mass estimate and the recent thermochemical model by \citet{trapman_exoalma_2025} -- which fits CO isotopologues and ${\rm N_2H^+}$ -- further supports that optically thin $\coo$ emission alone can yield reliable gas mass estimates when interpreted with self-consistent models \citep[see also][for a comparison across more Lupus disks]{Deng_2025_AGEPRO_III_Lupus}. 
This is particularly important for extending gas mass measurements to fainter disks, such as those around older or very low-mass stars, where ${\rm N_2H^+}$ is unlikely to be detectable within reasonable ALMA integration times. 
Overall, our finding that IM~Lup is a massive gas disk with no significant global CO depletion is in line with simple expectations for Myr-old disks \citep[see also][for an extension to other similarly large and aged disks]{pascucci_noCOdep_2023}. 
The new features of \texttt{DiskMINT}, which enable the separate fitting of dust and gas distributions, open the door to empirical estimates of the radial dust-to-gas mass ratio, a key parameter for assessing, for example, the potential for planetesimal formation. 
While IM~Lup appears to be a disk where the conditions for planetesimal formation are not yet met, many other disks with prominent dust ring structures may be more favorable. 
The method developed here for IM~Lup offers a promising path to constraining the planet-forming potential of such structured disks. 

\paragraph{Acknowledgments}

The authors thank C.P. Dullemond and A. Juhasz for their helpful assistance and suggestions in developing our code based on \texttt{RADMC-3D} and \texttt{radmc3dPy}.
The authors thank the anonymous referee, and also thank C. Pittman, A. Sierra, R. Teague, and A. Youdin for helpful suggestions and comments.
D.D., U.G., and I.P. acknowledge support from the NASA/XRP research grant 80NSSC20K0273.
D.D. and I.P. also acknowledge support from the Collaborative NSF Astronomy \& Astrophysics Research grant (ID: 2205870). 
I.P. also acknowledges partial support by the National Aeronautics and Space Administration under agreement No. 80NSSC21K0593 for the program ``Alien Earths''.

This work made use of the High Performance Computing (HPC) resources, which are supported by the University of Arizona TRIF, UITS, and Research, Innovation, and Impact (RII) and maintained by the University of Arizona Research Technologies department.
This paper makes use of the following ALMA data: ADS/JAO. ALMA\#2016.1.00484.L, ADS/JAO.ALMA\#2018.1.01055.L. ALMA is a partnership of ESO (representing its member states), NSF (USA) and NINS (Japan), together with NRC (Canada), NSTC and ASIAA (Taiwan), and KASI (Republic of Korea), in cooperation with the Republic of Chile. The Joint ALMA Observatory is operated by ESO, AUI/NRAO and NAOJ. The National Radio Astronomy Observatory is a facility of the National Science Foundation operated under cooperative agreement by Associated Universities, Inc.

\textit{Facility:} ALMA.

\textit{Software:}
All figures were generated with the \texttt{Python}-based package \texttt{MATPLOTLIB} \citep{Hunter2007}.
This research made use of \texttt{RADMC-3D} \citep[][]{Dullemond_radmc-3d_2012}, \texttt{LIME} \citep[][]{BrichHogerheijde_LIME_2010}, \texttt{Optool} \citep[][]{dominik_optool_2021}, \texttt{GoFish} \citep[][]{teague_gofish_2019}, \texttt{Astropy} \citep{astropy:2018}, and \texttt{Scipy} \citep{2020SciPy-NMeth}.
The models in this work is created with \texttt{DiskMINT} \citep[][]{Deng_2023_diskmint} v1.5.0 (frozen version released on \texttt{zenodo} \citealt{Deng_2025_diskmint_zenodo_v1p5}), and it can also be downloaded from our public \texttt{GitHub} repository: \url{https://github.com/DingshanDeng/DiskMINT}.

\newpage

\appendix
\restartappendixnumbering 
\twocolumngrid             

\section{Retrieving surface densities for the structured model via fits to observational data}
\label{appendix:model_A_to_B}

\begin{figure}[ht]
\gridline{\fig{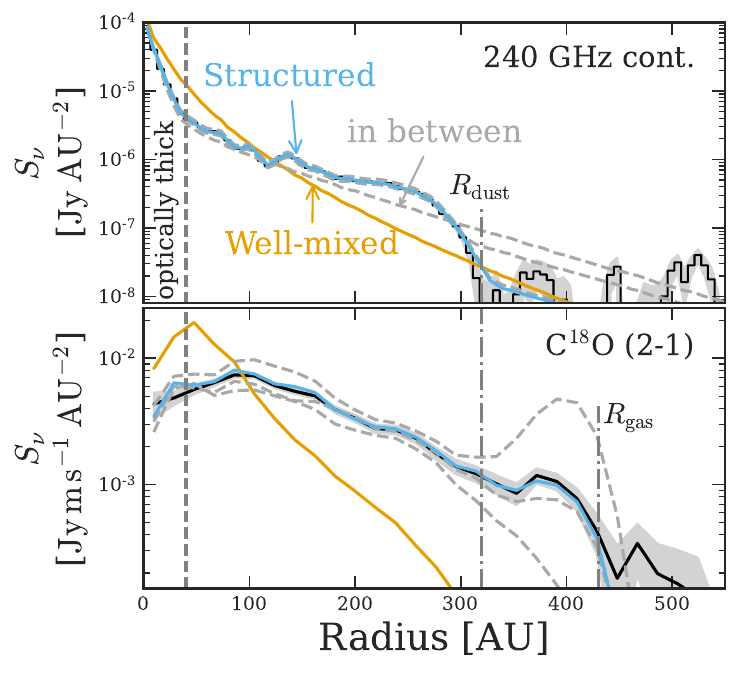}{0.47\textwidth}{}}
\caption{The radial profiles of dust continuum (top) and $\coo$ line emission (bottom) for well-mixed (orange lines) and structured (blue lines) models, and the intermediate ones (gray dashed lines) in between. The dust and gas outer radii are also shown in two vertical dot-dashed gray lines, and the vertical dashed gray line shows where the observation is getting optically thick in the inner disk. The notation follows Figure~\ref{fig:Result_Model_B}.}
\label{fig:Model_A_to_B}
\end{figure}

Here we present the radial profiles of the dust continuum and $\coo$ line emission of the well-mixed and structured models, and also show intermediate steps before final convergence in Figure~\ref{fig:Model_A_to_B}.
The well-mixed model uses a self-similar surface density distribution as described in Equation~\ref{eq:power-law-sigma} with constant $\varepsilon$, which is a reasonable approximation when there are no resolved observations of a disk.
The structured model  leverages additional spatial information when available by considering decoupled dust and gas densities, and retrieves both $\sigmadust(r)$ and $\sigmagas(r)$ by fitting the radial profiles of dust continuum and $\coo$ line emission through iterations (Equations~\ref{eq:fit_sigmadust}, \ref{eq:fit_sigmagas}). 
Some results from intermediate models before converging to the final solution are shown as gray dashed lines here. 
For more details on how the structured model is set up, see Section~\ref{subsec:data_driven_model_B}.

\section{Effect of dust settling}
\label{appendix:settling_models}

\begin{figure}[ht]
\gridline{\fig{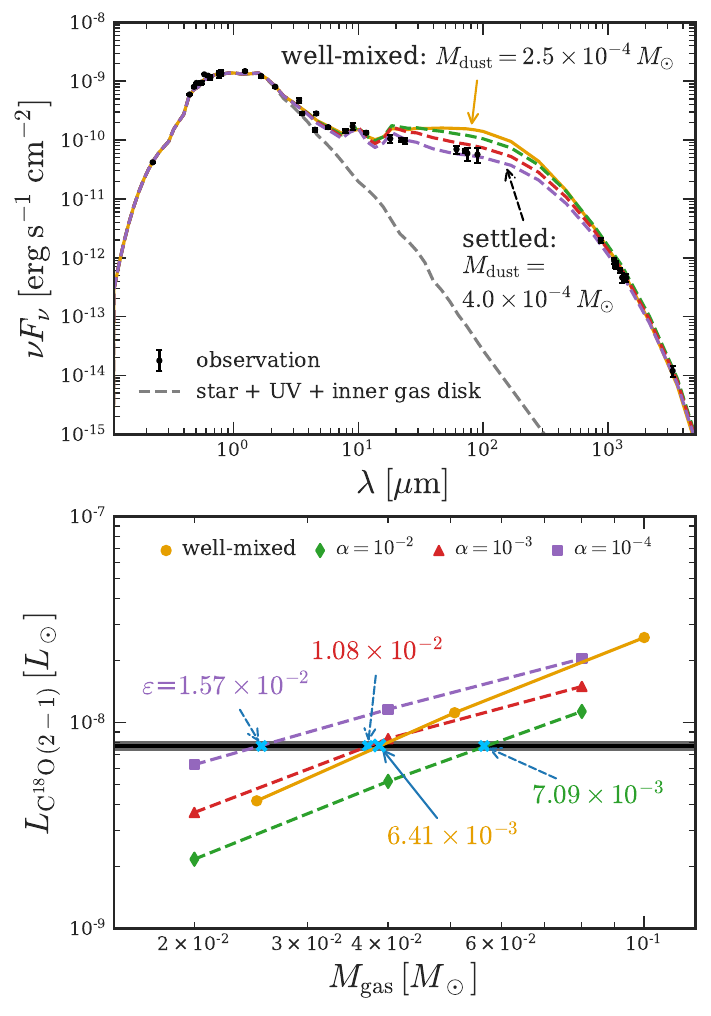}{0.45\textwidth}{}}
\caption{The SED fitting (top) and the $L_{\coo(2-1)}$ fitting (bottom) results for the well-mixed model comparing with the models that is settled. The figures are following the same notations as Figure~\ref{fig:Result_Model_A}. The blue line shows the results for the well-mixed model, with the well-mixed dust grains, while the green, blue and purple lines (and scatters) are three models with dust settling, which has $\alpha_{\rm v} = 10^{-4}, 10^{-3}, 10^{-2}$, respectively. Their best-fit $\mdust$ are shown in the top panel. In the lower panel, three different grid points with $\varepsilon = 0.02, 0.01, 0.005$ are shown as scatter points, with their best-fit $\varepsilon$ indicated by blue crosses, and with their values annotated with arrows. The black horizontal line represents the observed $L_{\coo(2-1)}$ with the gray area indicating its uncertainty.}
\label{fig:Result_Model_A_dustset}
\end{figure}

In this work, we newly introduced dust settling into \texttt{DiskMINT} as we described in Section~\ref{subsec:data_driven_model_B}.
While dust settling has a negligible role in the thermal balance \citep[][]{Ruaud2022}, it could have some impact on the derived dust properties, particularly dust mass estimates.
Here, we run \texttt{DiskMINT} with the same setup as well-mixed model -- so that the surface density is described as Equations~\ref{eq:power-law-sigma} with $r_c = 100\,{\rm AU}$ and $\gamma = 1.0$ -- but apply dust settling with three different $\alpha_{\rm v} = 10^{-4}, 10^{-3}, 10^{-2}$, and present the results in Figure~\ref{fig:Result_Model_A_dustset}.
We find that the models with dust settling (green, red, and purple lines) match the SED better at $\lambda \sim 100\,\um$, while the well-mixed model (blue) overproduces the emission there.
We also find the models with smaller $\alpha_{\rm v}$ result in a higher flux at $\sim 100\,\um$, but there is no obvious difference between the models with settling at longer wavelengths mm-wavelengths.
Because dust settling also lowers the SED at longer wavelengths ($\lambda \gtrsim 100\,\um$), the $\mdust$ is increased to $4.0\times10^{-4}\,M_\odot$ to reproduce the SED at those long wavelengths.

We also ran the models with $\varepsilon = 0.02, 0.01, 0.005$ and get the synthesized $L_{\coo(2-1)}$ and compare with the observations.
We find the best-fit $\varepsilon = 0.007-0.016$ with $\mgas = 0.025 - 0.056\,M_\odot$, which is very close to the derived $\mgas = 0.039\,M_\odot$ from the well-mixed model. 
We also find that neither the $\mdust$ nor $\mgas$ estimates are sensitive to the choice of $\alpha_{\rm v}$, and therefore we conclude that dust settling does not have a significant impact on disk mass estimates.

\section{Dust fragmentation and drift}
\label{appendix:model_C_theories}

\subsection{Dust fragmentation}
As dust particles collide in the disk, if their relative velocity exceeds the threshold fragmentation velocity, they fragment rather than stick and grow.
Larger grains tend to collide at higher impact velocities, and they are more likely to shatter.
This process sets the upper limit of the grain size at different locations in disks.
We consider the dust fragmentation by including the threshold grain size following \citet{birnstiel_simple_2012}, 
\begin{equation}\label{equation:fragmentation_size}
    a_{\rm frag} = \frac{2}{3\pi} \frac{u_{\rm frag}^2}{\rho_\bullet \alpha_{\rm v} c_{s, {\rm mid}}^2} \sigmagas,
\end{equation}
where $u_{\rm frag}$ is the fragmentation threshold velocity, $\alpha_{\rm v}$ is the turbulent parameter, and the $c_{s, {\rm mid}}$ here is the sound speed at disk mid-plane ($c_{s, {\rm mid}} = k_B T_{{\rm gas},{\rm mid}}/\mu$). 
Here $u_{\rm frag}$ and $\alpha_{\rm v}$ are constrained from the observations.
We assume that all the grains with size $a \geq a_{\rm frag}$ are fully fragmented, and their masses will be distributed to all the smaller grains with $a < a_{\rm frag}$, 
\begin{equation}
    \sigmadust^\prime(r, a) = 
    \left\{
    \begin{array}{ll}
      0;   &  (a \geq a_{\rm frag})\\
    \sigmadust^{\rm o}(r, a)\,{\rm ME_{frag}}(r); & (a < a_{\rm frag})
    \end{array}
    \right.
\end{equation}
where the $\sigmadust^\prime$ is the new dust surface density for each grain at each radius after we consider the fragmentation ${\rm ME_{frag}}(r)$ is the mass enhancement factor, which is the ratio between the total dust surface density and the original dust surface density ($\sigmadust^{\rm o}(r,a)$) for the smaller grains that are not fragmented, 
\begin{equation}
   {\rm ME_{frag}}(r) = \frac{\int_{a_{\rm min}}^{ a_{\rm max}} \sigmadust^{\rm o}(r, a)}{\int_{a_{\rm min}}^{ a_{\rm frag}} \sigmadust^{\rm o}(r, a )}.
\end{equation}
This ensures that we retain and redistribute the mass in the fragmented grains at the same radius. 
The overall dust mass distribution combining all grains does not change after considering the fragmentation $\int_a \sigmadust^{\prime}(r, a) = \sigmadust^{\rm o}(r) = \sigmagas(r)\varepsilon(r)$. 
However, the $\sigmadust^{\prime}(r, a)$ for different grain sizes are different from the original $\sigmadust^{\rm o}(r,a)$.
Because the millimeter-wavelength continuum emission is mostly the emission from the millimeter-sized grains ($a_{\rm mm}$), so that the $\sigmadust^{\prime}(r,a_{\rm mm})$ is inferred from the observation, which can further constrain the free-parameter $u_{\rm frag}$.

\subsection{Dust radial drifting}

Besides fragmentation, radial drift could also be important in the dust disk evolution and effectively reduce the disk size.
As the dust moves at Keplerian velocity while the gas moves at sub-Keplerian velocity, dust particles feel a headwind; thus, they lose angular momentum and drift inward.
This effect is sensitive to the particle size \citep[e.g.,][]{whipple_certain_1972}, resulting in different radial distributions for different particle sizes $\sigmadust(r, a)$.

Here we estimate the radial drift from the observations, by assuming that the dust disk boundary ($R_{\rm dust}$) is shaped by the radial drift barrier.
We introduce two free parameters: drifting threshold size $a_{\rm drift}$ and drifting radius $r_{\rm drift}$.
We move all the mass of large dust grains with $a \geq a_{\rm drift}$ inside $r_{\rm drift}$, and then distribute that mass as follows,

\begin{equation}
\begin{aligned}
\Sigma^{\prime\prime}_{\text{dust}}&(r, a) = \\
&\left\{
    \begin{array}{ll}
    \sigmadust^{\prime}(r, a); & \quad (r > r_{\text{drift}} \text{ and } a < a_{\text{drift}}) \\
    0; & \quad (r > r_{\text{drift}} \text{ and } a \geq a_{\text{drift}}) \\
    \sigmadust^{\prime}(r, a) \times {\rm ME_{drift}}; & \quad (r \leq r_{\text{drift}})
    \end{array}
\right.
\end{aligned}
\end{equation}

where the $\sigmadust^{\prime\prime}$ is the dust surface density considering the radial drift. For the grains that are outside of the defined drift radius $r > r_{\rm drift}$, the smaller grains ($a < a_{\rm drift}$) are not affected in this process, and the larger ones ($a \geq a_{\rm drift}$) are drifted in so that there will be zero mass left beyond $r_{\rm drift}$. 
For the grains inside $r < r_{\rm drift}$, their surface density is enhanced by the mass of the large grains that are drifted in. The $\sigmadust(r,a)$ inside $r_{\rm drift}$ are multiplied by a mass enhancement factor ${\rm ME_{drift}}$,

\begin{equation}
{\rm ME_{drift}} = \frac{\int_{r_{\rm in}}^{ r_{\rm drift}} \int_{a_{\rm min}}^{ a_{\rm max}} \sigmadust^{\prime}(r, a) + \int_{r_{\rm drift}}^{ r_{\rm out}} \int_{a_{\rm drift}}^{ a_{\rm max}} \sigmadust^{\prime}(r, a)}{\int_{r_{\rm in}}^{ r_{\rm drift}} \int_{a_{\rm min}}^{ a_{\rm max}} \sigmadust^{\prime}(r, a )}.
\end{equation}

which is the ratio of the total mass of all grains inside $r_{\rm drift}$ plus the drifted-in large grains beyond $r_{\rm drift}$, versus that original mass inside $r_{\rm drift}$. 
This ensures that the radial drift removed the large grains with $a \geq a_{\rm drift}$ at $r > r_{\rm drift}$, but it does not change the grain size distribution and their relative mass fractions for each particle inside the $r_{\rm drift}$.

The drift threshold size $a_{\rm drift}$ can also be estimated as \citep[e.g., ][]{birnstiel_simple_2012, birnstiel_dust_2024}:
\begin{equation}\label{equation:amax_drift}
    a_{\rm drift}(r) = \frac{2 \sigmadust}{\pi \rho_\bullet \gamma} \left(\frac{h}{r}\right)^{-2},
\end{equation}
where the $\rho_{\bullet}$ is the bulk grain density that is a constant everywhere, and is calculated from the dust composition according to the mass fractions, and $h$ is the gas pressure scale height evaluated at the midplane $(z=0)$, 
\begin{equation}
    h(r, z) = \sqrt{\frac{k_B}{G M_\star \mu} T_{\rm gas} (r^2 + z^2)^{\frac{3}{2}}},
\end{equation}
where $k_B$ is the Boltzmann constant, $G$ is the gravitational constant, $M_\star$ is the stellar mass, $\mu$ is the mean molecular weight.

\section{Band~3 analysis} \label{appendix:model_B_band3}

\begin{figure}[t]
\gridline{\fig{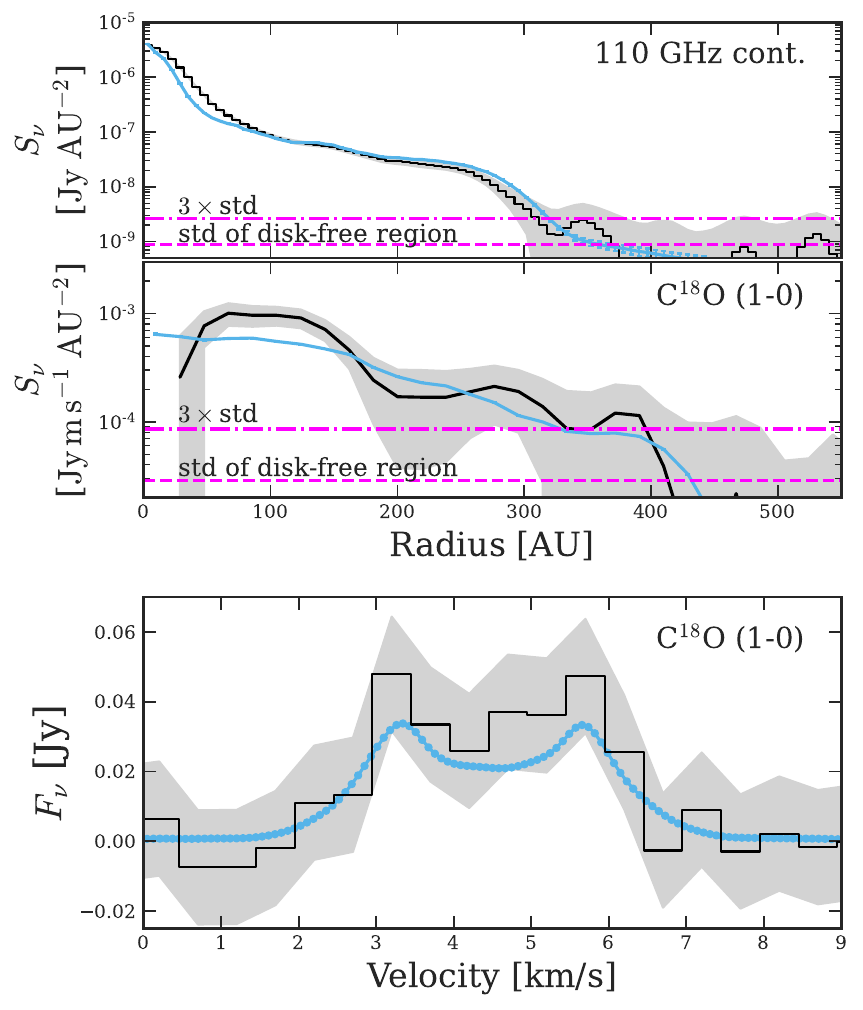}{0.47\textwidth}{}}
\caption{The results for the structured model compared with the MAPS observations in ALMA Band~3. The top panel shows the comparison with the dust continuum emission at 110 GHz \citep[][]{sierra_molecules_2021}, and the middle and bottom panels show the comparison with the $\coo\,$(1-0) line \citep[][]{zhang_MAPS_molecules_2021}. The notation follows Figure~\ref{fig:Result_Model_B}.}
\label{fig:Model_B_Band3}
\end{figure}

Here we present the results of the ALMA Band~3 continuum at 110 GHz and $\coo\,(1-0)$ line emission for the structured model in Figure~\ref{fig:Model_B_Band3}.
Note that the models are fit to the higher quality ALMA Band~6 data, without considering the poorer quality Band~3 information. 
Both the continuum and $\coo\,(1-0)$ line emission from the best-fit model match the these longer wavelength data within the uncertainties.
The continuum emission shows that the slightly larger dust grains ($\sim 0.03\,{\rm cm}$) traced at 110 GHz have a similar distribution compared to the grains ($\sim 0.01\,{\rm cm}$) traced at 240 GHz. 
The $\coo\,{1-0}$ line emission, coming from a colder region that is closer to the disk midplane, shows that our model provides good estimates on the CO snowline and supports the CO$_2$/CO conversion as a cause for lower $\coo$ emission in disks rather than a homogeneous reduction in CO abundance everywhere.



\bibliography{IMLup}{}
\bibliographystyle{aasjournalv7}

\end{document}